\documentclass{aa}
\usepackage{amssymb}
\usepackage{amsmath}
\usepackage{txfonts}
\usepackage{graphicx}
\usepackage{natbib}
\usepackage{url}

\bibpunct{(}{)}{;}{a}{}{,} % to follow aa

\newcommand{\lhls}{\ensuremath{\ell_\text{h}/\ell_\text{s}}}

\begin{document}

\title{Long Term Variability of Cyg~X-1}
\subtitle{IV. Spectral Evolution 1999--2004}

\author{J.~Wilms\inst{1} \and M.A.~Nowak\inst{2} \and
   K.~Pottschmidt\inst{3} \and G.G.~Pooley\inst{4} \and S.~Fritz\inst{5}}
\institute{
  Department of Physics, University of Warwick, Coventry, CV7 1AL, UK
\and MIT-CXC, NE80-6077, 77 Massachusetts Ave., Cambridge, MA 02139, USA
\and Center for Astrophysics and Space Sciences, University of
   California at San Diego, La Jolla, CA 92093-0424, USA
\and Mullard Radio Astronomy Observatory, Cavendish Laboratory, Madingley
  Road, Cambridge CB3 0HE, UK 
\and Institut f\"ur Astronomie und Astrophysik, Universit\"at
   T\"ubingen, Sand 1, 72076 T\"ubingen, Germany
}

\mail{J.~Wilms (j.wilms@warwick.ac.uk)}
\titlerunning{Long Term Variability of Cygnus~X-1: IV.}
\authorrunning{J.~Wilms et al.}
\date{Received $<$date$>$ / Accepted $<$date$>$ }

\abstract{Continuing the observational campaign initiated by our
  group, we present the long term spectral evolution of the Galactic
  black hole candidate Cygnus~X-1 in the X-rays and at 15\,GHz.  We
  present $\sim$200 pointed observations taken between early 1999 and
  late 2004 with the Rossi X-ray Timing Explorer and the Ryle radio
  telescope.  The X-ray spectra are remarkably well described by a
  simple broken power law spectrum with an exponential cutoff.
  Physically motivated Comptonization models, e.g., by
  \citet[\texttt{compTT}]{tit:94a} and by \citet[eqpair]{coppi:99a},
  can reproduce this simplicity; however, the success of the
  phenomenological broken power law models cautions against
  ``overparameterizing'' the more physical models.  Broken power law
  models reveal a significant linear correlation between the photon
  index of the lower energy power law and the hardening of the power
  law at $\sim$10\,keV.  This phenomenological soft/hard power law
  correlation is partly attributable to correlations of broad band
  continuum components, rather than being dominated by the weak
  hardness/reflection fraction correlation present in the
  Comptonization model.  Specifically, the Comptonization models show
  that the bolometric flux of a soft excess (e.g., disk component) is
  strongly correlated with the compactness ratio of the Comptonizing
  medium, with $L_\text{disk} \propto
  (\ell_\text{h}/\ell_\text{s})^{-0.19}$.  Over the course of our
  campaign, Cyg~X-1 transited several times into the soft state, and
  exhibited a large number of ``failed state transitions''.  The
  fraction of the time spent in such low radio emission/soft X-ray
  spectral states has increased from $\sim$10\% in 1996--2000 to
  $\sim$34\% since early 2000.  We find that radio flares typically
  occur during state transitions and failed state transitions (at
  $\ell_\text{h}/\ell_\text{s}\sim 3$), and that there is a strong
  correlation between the 10--50\,keV X-ray flux and the radio
  luminosity of the source.  We demonstrate that rather than there
  being distinctly separated states, in contrast to the timing
  properties the spectrum of Cyg X-1 shows variations between extremes
  of properties, with clear cut examples of spectra at every
  intermediate point in the observed spectral correlations.
  }

\maketitle

\section{Introduction}\label{sec:intro}
One of the major results of X-ray and $\gamma$-ray astronomy since the
discovery of Galactic black hole candidates (BHCs) more than thirty
years ago has been the realization of the complexity of the spectral
variability of these sources on time scales from milliseconds to
decades. It is generally believed that this variability can provide
clues as to the basic physical processes in X-ray
binaries. Importantly, the broad-band spectra and timing behavior of
accreting stellar mass black holes in X-ray binaries and of
supermassive black holes in the centers of Active Galactic Nuclei
(AGN) are very similar \citep{uttley:02a,uttley:05a}. Due to the
scaling of the dynamical time scales in these systems with the mass of
the central object, studying Galactic and extragalactic systems is
complementary. Galactic sources allow us to see how the
accretion process evolves on very long time scales, while AGN provide
``snapshot'' observations of the accretion process, over
characteristic dynamical time scales, with a signal to noise ratio
that is not reachable for Galactic sources.

Observations over the last three decades have shown that Galactic BHC
exhibit distinct and very characteristic states with different
spectral shape and variability behavior, with AGN seemingly
following similar trends \citep[][and
therein]{mchardy:05a,jester:05a}.  The state of a BHC depends on not
yet fully understood parameters \citep{homan:01a}, although it is
believed that the mass accretion rate, $\dot{M}$, and therefore the
luminosity of the accreting system, plays a vital role.  See
\citet{mcclintock:05a} and \citet{belloni:04a} for an extensive
discussion of these issues.

For low accretion luminosities in the ``hard state'' of Galactic black
holes, the X-ray spectrum can be described by a hard power law with
photon index $\Gamma\sim 1.7$ and an exponential cutoff at
$\sim$150\,keV. At low energies, most BHC show some kind of soft
excess with a characteristic temperature of a few 100\,eV. This
baseline continuum is modified by a Fe K$\alpha$ emission line at
$\sim$6.4\,keV and by reflection features, indicating the close
proximity of the source of hard X-rays and relatively cold material.
The hard state is furthermore characterized by strong X-ray
variability of $>$10\% rms and by the presence of radio emission.  In
systems where the radio sources have been resolved, this radio
emission has been shown to originate in an outflow that is consistent
with being mildly relativistic ($v\sim 0.6c$) in the hard state
\citep{stirling:01a,gallo:03a}. For Cyg~X-1, this outflow has recently
been shown to have a time averaged kinetic power of
$\sim$$10^{36}\,\text{erg}\,\text{s}^{-1}$ to
$10^{37}\,\text{erg}\,\text{s}^{-1}$, i.e., a significant fraction of
the system's X-ray luminosity \citep{gallo:05a}.

In contrast to the ``hard state'', the (typically) higher luminosity ``soft
state'' exhibits a soft X-ray spectrum that can be well described by
thermal emission from a standard accretion disk. Where a hard spectral
component is detected, it does not show any appreciable curvature,
sometimes up to the MeV regime
\citep{grove:98a,mcconnell:00a,mcconnell:01a}. The X-ray variability
during the soft state is weak ($\lesssim 6$\% rms) and no radio
emission is detected \citep{fender:99b}.

While there is general agreement that the soft excess is due to
emission from an accretion disk, the interpretation of the hard
spectral component and its relationship to the reprocessing features
is still debated. The canonical interpretation of the hard state X-ray
spectrum, first proposed in the 1970s
  \citep[e.g.,][]{thorne:75a,shapiro:76a,sunyaev:79a} and later
elaborated upon, e.g., by \citet{haardt:91a} and \citet{haardt:93a},
is that the hard spectral component is caused by Comptonization, where
soft X-rays from the inner disk are Compton upscattered by hot
electrons ($kT_\text{e}\sim 100$\,keV) in a predominantly thermal
electron gas, often called the ``accretion disk corona'' (ADC). The
spectrum emerging from the corona has the proper power law plus
exponential cutoff shape.  To explain the presence of the ADC,
magnetohydrodynamical instabilities \citep[e.g.,][and references
therein]{balbus:98a}, which only work efficiently at lower $\dot{M}$,
have been invoked.

Observations limit the covering factor of the ADC with respect to the
source of soft X-rays to $\ll$1. Otherwise the ADC would fail to reach
the high coronal temperatures, inferred from the energy of the
exponential cutoff, due to efficient Compton cooling \citep[e.g.,][and
therein]{haardt:97a,dove:97a,dove:97b}.  Geometric models invoked to
achieve smaller ADC covering factors include patchy coronae
\citep{stern:95b}, magnetic flares
\citep{beloborodov:99a,poutanen:99b}, advection dominated
accretion flows \citep{esin:98a}, ADCs with non-static or outflowing
coronae \citep{beloborodov:00a,malzac:01a}, and accretion
flows with an inner geometrically thick ADC and an outer geometrically
thin and optically thick accretion disk \citep{dove:97a,dove:97b}. In
all of these models, a fraction of the hard X-rays is scattered back
towards the disk, giving rise to fluorescence Fe K$\alpha$ emission
and the Compton reflection hump \citep{lightman:79a,lightman:88a}.
Comparisons between observations and theoretical model spectra for all
of these different geometries have been successful.

With the realization of the importance of jets and their association
with the hard state, the ``Comptonization paradigm'' has recently been
challenged. The discovery of a tight correlation between the radio and
the X-ray emission on time scales of days
\citep{hannikainen:98a,pooley:98a,corbel:00a} suggests a close
coupling between the X-ray and the radio emitting media
\citep{markoff:02a,heinz:03a}. In addition, in a direct comparison
between data and theory, jet models capable of reproducing the
radio--X-ray correlations have been recently shown to describe the
X-ray spectrum with a precision comparable to Comptonization models
\citep{markoff:05a}. The origin of the X-rays in these models is due
to a combination of synchrotron emission from the jet and synchrotron
self Compton emission from the jet base. We note, however, that a
small separation of the emission regions cannot be ruled out from
joint radio--X-ray timing timing arguments \citep{gleissner:04a},
furthermore, accretion models postulating a strong coupling between a
Comptonizing medium and the jet are also able to explain the
radio--X-ray correlation without postulating a jet origin for the
X-rays \citep{meier:01a,merloni:02a}.

In conclusion, all existing models suggest a complex interplay between
the energetics and emission from the accretion disk, Comptonizing
plasma, and radio jet. This interplay can be disturbed by changes in
some external parameter, such as $\dot{M}$. These changes may lead to
state changes as well as to subtle changes in the overall source
properties, such as the so called ``failed state transitions''
\citep{pottschmidt:00b}.  An empirical understanding of the overall
accretion process is therefore difficult to gain from only a few
single and isolated observations, although these are very important
for determining broad band spectra or for measuring spectra with high
energy resolution \citep[e.g.,][]{miller:02a}. Monitoring campaigns
covering the characteristic time scales of these spectral changes are
required. We note that the necessity of monitoring campaigns follows
also from hysteresis effects seen in many sources, i.e., the earlier
source history is important for determining the source properties in a
given observation \citep{miyamoto:95a,nowak:02a,maccarone:02b}.

Monitoring campaigns using large effective area and proportional
counter energy resolution became possible in the 1990s with the
\textsl{Rossi X-ray Timing Explorer} (\textsl{RXTE}).  Since then,
monitoring campaigns using \textsl{RXTE}'s pointed instruments, the
Proportional Counter Array, \citep[PCA;][]{jahoda:96b} and the High
Energy X-ray Timing Experiment, \citep[HEXTE;][]{rothschild:98a}, have
been performed for virtually all known persistent and transient
Galactic black hole candidates.  Among many others, examples for such
campaigns are those on \object{LMC X-1} and \object{LMC X-3}
\citep{nowak:99f,wilms:99d}, \object{GX~339$-$4}
\citep{wilms:98c,nowak:99a,nowak:02a,kong:02a,homan:05a,belloni:05a},
\object{V1408~Aql} \citep{nowak:99b}, \object{4U~1543$-$47}
\citep{kalemci:05a}, \object{XTE J1650$-$500} \citep{kalemci:03a},
\object{4U~1630$-$47} \citep{tomsick:00a}, and \object{XTE J1550$-$564}
  \citep{kalemci:01a,sobczak:00a}.

In this paper, we present results from one of the longest of these
monitoring campaigns, which was initiated by the authors in 1996 to be
performed on the prototypical hard state black hole candidate (BHC)
\object{Cygnus X-1}.  Our campaign consists of biweekly, $\sim$5\,ksec
long \textsl{RXTE} pointings and simultaneous radio observations at
15\,GHz ($\lambda=2\,\text{cm}$) with the Mullard Radio Astronomy
Observatory's Ryle telescope in Cambridge, UK.  Earlier papers in this
series concentrated on the evolution of X-ray time lags
\citep{pottschmidt:00b}, of the power spectrum \citep[][hereafter
paper~\textsc{i}]{pottschmidt:02a}, the linear relationship between
the short term root mean square variability and the flux
\citep[][hereafter paper~\textsc{ii}]{gleissner:03a}, and correlations
between the soft X-rays and the radio flux \citep[][hereafter
paper~\textsc{iii}]{gleissner:04a}.  Amongst others, archival data
from the campaign have also been used in the discovery of giant flares
from Cyg~X-1 \citep{gierlinski:03a}, a subset of the archival
observations was also used in studies of Cyg X-1's spectral and
temporal variability \citep{axelsson:05a,ibragimov:05a}. Here, we
discuss the results obtained on the spectral evolution of Cyg~X-1
throughout the \textsl{RXTE} monitoring campaign, starting in
Sect.~\ref{sec:reduction} with a description of the changes of our
data reduction procedure with respect to papers~\textsc{i}
through~\textsc{iii}.  In Sect.~\ref{sec:evolution} we give the
results of modeling the broad band 3--200\,keV X-ray spectra using
empirical spectral models as well as advanced Comptonization models
and consider the evidence for interaction between the X-ray and the
radio emission.  We summarize our results in Sect.~\ref{sec:summary}.
We will discuss the joint spectral-timing behavior of Cyg~X-1 based on
our analyses in a future paper.

\section{Data Reduction}\label{sec:reduction}
\subsection{Overview}

We have already given an extensive overview of the \textsl{RXTE}
observing strategy and our data extraction in papers~\textsc{i}
through \textsc{iii}.  Here, we only highlight the most important
points pertaining to the X-ray spectral analysis, mainly centered on
the RXTE calibration.

In this paper we consider 202 observations with a typical good time of
3--7\,ksec, for a total good time of $\sim$989\,ksec.  We use data
from both the PCA and the HEXTE.  Spectra are extracted with
HEASOFT~5.3.1 and then fit with XSPEC 11.3.1w \citep{arnaud:96a}.  For
the spectral analysis, we generated 3--25\,keV PCA spectra from the
top Xenon layer data and analyze spectra taken with different numbers
of proportional counter units (PCUs) separately. This choice was made
to allow the comparability with the results of the timing analysis.
For the HEXTE, data from 18--120\,keV were considered.

As shown in Appendix~\ref{app:speccal}, compared to earlier versions
of the response matrices, the PCA is now in much better agreement with
the HEXTE and also with other missions, in both, flux and spectral
slope, with the remaining uncertainty being taken into account by a
multiplicative constant that is normally very close to unity.  Our
reevaluation of the PCA calibration also shows that remaining
calibration uncertainties can be modeled to first order by adding a
systematic uncertainty of 0.5\% to the data, although even with this
choice some significant calibration effects are still visible in the
residuals (see Appendix~\ref{app:speccal}).  As shown in
Appendix~\ref{app:systematics}, however, this systematic error leads
to a dramatic overestimation of the range of the confidence
intervals for the fit parameters. Thus, although we compute confidence
intervals at the 90\% level for one interesting parameter, we do not
show error bars in the figures.

The good time of the observations is mainly defined by excluding times
of high PCA background. Due to the requirements of X-ray timing
analysis, early in the campaign very conservative constraints were
chosen, i.e., data taken within 30\,minutes of passages through the
South Atlantic Anomaly (SAA) and during times of high particle
background were discarded.  While in principle improved PCA background
models now allow for good results even closer to the SAA passages, for
consistency reasons and in order to facilitate the comparison of
spectral parameters to the results of papers~\textsc{i}
through~\textsc{iii}, the previous limits were retained.

Electronic tables in the Flexible Image Transport System (FITS) format
containing the results of the spectral fitting and the confidence
intervals for all best fit parameters are available accompanying the
electronic version of this paper\footnote{\label{fn:aatab}Tables~1--3
  are only available in electronic form at the CDS via anonymous ftp
  to cdsarc.u-strasbg.fr (130.79.128.5) or via
  \url{http://cdsweb.u-strasbg.fr/cgi-bin/qcat?J/A+A/}}. In our
discussion of peculiar observations below, we identify observations in
a way that makes it easy to find them in these FITS files and in the
RXTE archive. The abbreviated syntax used is of the type P$xxxxx$/$XX$
where $xxxxx$ is the RXTE proposal ID and where $XX$ is the number of
the observation within each proposal. We also give the date and time
of the observation to the closest hour and, in cases where an
observation resulted in more than one spectrum because different
detector combinations were used, we identify those detectors that were
off.

\subsection{Fitting Strategy}
For larger data sets, the CPU time needed to perform spectral fitting
can be considerable and therefore the choice of starting parameters
for spectral modeling and how the spectral fits are performed is of
some practical importance.  We note that the most obvious strategy of
spectral modeling, namely using the best fit values of nearby
observations, can severely bias the correlations found between
different spectral parameters as this strategy can force larger
numbers of data points onto local $\chi^2$-minima. This bias is less
of an issue when always using the same starting parameters, which is
the approach adopted for the fits described below, even though this
approach requires a larger amount of CPU time.

We also find that for complex models, such as the Comptonization fits
described in Sect.~\ref{sec:comptt} and~\ref{sec:eqpair}, the
$\chi^2$-valley exhibits a large number of local minima and there are
many degeneracies between parameter combinations.  Both effects result
in a significant fraction of fits converging on local $\chi^2$-minima.
Even when employing standard tricks, such as computing error contours,
the global minimum was not always reached.  These ``rogue fits'' can
be identified through outliers in one or more parameters when looking
at correlations between different spectral parameters. Where we
identified such bad fits we refit the data using a different set of
starting parameters obtained from ``typical fits'' using the same
model.

A potential drawback of this approach is that it effectively forces
these observations onto the correlations. Only about 20\% of our
observations, however, had to be treated in this labor intensive way.
For a small fraction of these refitted observations, the
$\chi^2_\text{red}$ of the finally adopted spectral model was found to
be \emph{larger} by ${\cal O}(0.01)$ than the original
$\chi^2_\text{red}$ due to degeneracies between the fitting
parameters. A typical example here are fits where the Fe K$\alpha$
line energy pegs at its (unphysical) lower bound of 6\,keV, which is
often the case because of the PCA calibration uncertainty around the
Xe~L-edge. In such cases we searched for better models with the Fe
line energy at $\sim$6.4\,keV and accepted the best fit parameters
from these fits, even if their $\chi^2$ was slightly higher. The
number of observations for which such a strategy was required is still
small enough, and the increase of $\chi^2$ was only ${\cal O}(0.01)$,
such that this strategy does not influence the results listed below
and we are confident that all parameters shown in the following
reflect the physical behavior of Cyg~X-1.

In general, for all spectral fits presented here we find that for the
reduced $\chi^2$, $\chi^2_\text{red}\lesssim 2$. Values of $\chi^2$
clearly larger than~1 might appear worrisome at first, however, we
stress that the signal to noise ratio of our observations is very high
and that the available spectral models are often too simple to be able
to describe all subtleties of the observations. Furthermore, given our
choice of systematics, for lower $\chi^2_\text{red}$ PCA data are
completely dominated by calibration uncertainties. In principle this
would require data analysis methods that treat systematic errors in a
much more elaborate way.  We stress that typical observations of
bright sources with other satellites, such as \textsl{Chandra} or
\textsl{XMM-Newton}, result in good $\chi^2$ values despite having
significantly \emph{larger} (10\% or more) deviations between the data
and the model than our \textsl{RXTE} fits, where we typically find
that data and model deviate by $<$1\%.  These better $\chi^2$ values
from \textsl{Chandra} and \textsl{XMM-Newton} are by virtue of their
smaller effective areas and consequently larger Poisson errors which
dominates over the calibration uncertainty of these satellites.  Seen
in this light, our \textsl{RXTE} models thus provide an overall good
description of the spectral shape of Cygnus X-1.

\begin{figure*}
\includegraphics[width=12cm]{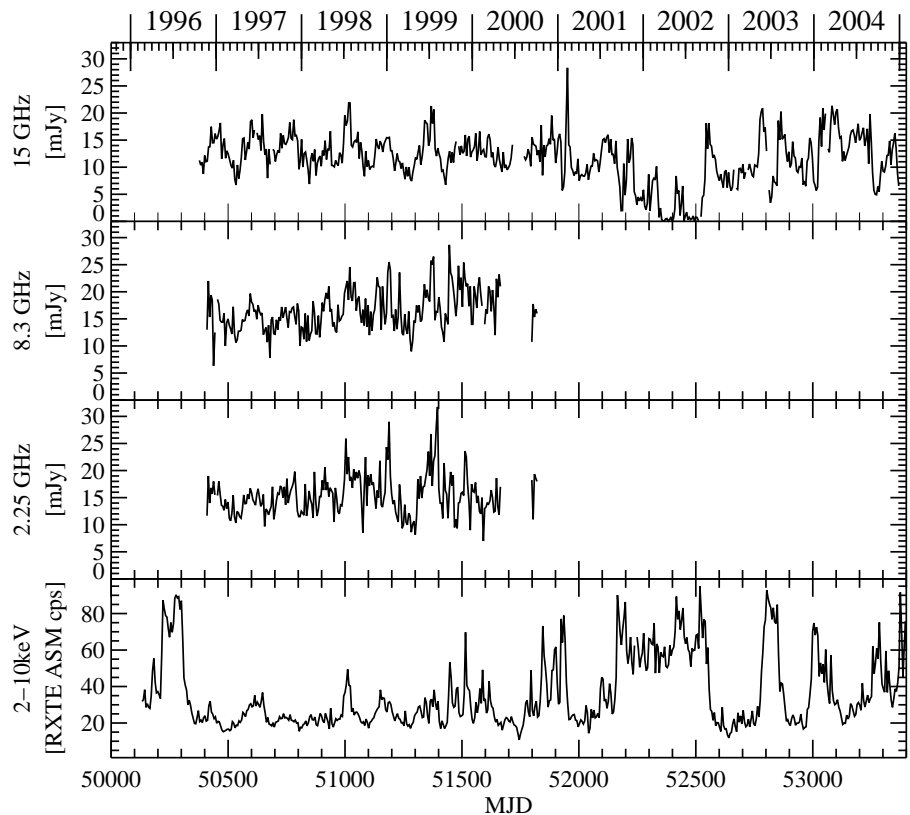}
\hfill\begin{minipage}[b]{5.5cm}
\caption{\textsl{RXTE}-ASM 2--10\,keV, and radio lightcurves of Cyg
  X-1 from 1996 until the end of 2004. The 15\,GHz data are from the
  Ryle telescope, the 8.3\,GHz and 2.25\,GHz measurements come from
  the National Radio Astronomy Observatory's Green Bank
  Interferometer. The data are rebinned to a resolution of 5.6\,d to
  smooth out the orbital variability.}\label{fig:evolution}
\end{minipage}
\end{figure*}

\section{Modeling the 3--200\,keV Spectrum of Cygnus
  X-1}\label{sec:evolution}

\subsection{Cygnus X-1 since the launch of \textsl{RXTE}}
Fig.~\ref{fig:evolution} shows the X-ray and radio lightcurves of
Cyg~X-1 for the nine year long interval from early 1996 (launch of
\textsl{RXTE}) until the end of 2004. Shown are the 2--10\,keV X-ray
flux measured with the \textsl{RXTE} All Sky Monitor
\citep{levine:96a}, as well as the radio lightcurves at 15\,GHz from
the Ryle telescope and at 2.25\,GHz and 8.3\,GHz measured with the
Green Bank interferometer (GBI; these data are only available until
early 2000 when the GBI monitoring campaign was discontinued due to
lack of funding). To minimize the significant source variability on its
orbital time scale \citep{brocksopp:99b} and to emphasize the
long-term evolution of the source, the data were rebinned to a
resolution of 5.6\,d.

The figure clearly shows the well known fact that Cyg~X-1 transits
between two states. As will be elaborated upon below, for ASM count
rates below $\sim$45\,cps, the source is in a classical hard state.
During this time radio emission is observed at a level of
$\sim$12\,mJy at 15\,GHz.  In contrast, during times when the ASM
count rate is $\gtrsim$80\,cps, the source spectrum is soft, although
a hard power law component is still seen. We will call these phases
the ``soft state'', although it is important to note that Cyg~X-1 only
rarely reaches what one would call a classical ``soft state'' in
transient BHC. During these soft states, the radio emission is
strongly reduced. We especially note that during the long soft state
of 2001/2002 there were phases where no radio emission could be
detected \citep[mm-emission was quenched during that time as
well;][]{tigelaar:04a}.

During transitions between the hard and soft states, radio flares are
observed where the radio flux can be up to a factor of two higher
than during the hard state.  Cyg~X-1 also shows short episodes where
the ASM count rate increases. We have identified these episodes with
``failed state transitions'' before (paper~\textsc{i}), since they are
characterized by the X-ray spectrum softening and by clearly changed
timing properties \citep[e.g., increased X-ray time
lags][]{pottschmidt:00b}, similar to state transitions, although the
source never really settles into a soft state like behavior.

Prior to $\sim$1999, Cyg~X-1 was in the hard state for most of the
time, except for rather short ``failed state transitions'' and the
short soft state of 1996. From that time onwards, the frequency of the
failed state transitions increased until, from 2001 July onwards,
Cyg~X-1 reached the long 2001/2002 soft state. It stayed in this phase
for almost a year until 2002 August/September where it rapidly
transited back into the hard state, with a spectrum similar to before
the soft state. Shortly after this, in 2003 June and at the turn of
2003/2004, two other short soft state episodes occured. From early
2004 onwards, the \textsl{RXTE} ASM count rate was generally increased
with respect to the 1997--1999 hard states and the source spectrum was
generally softer as well.

We can quantify this behavior for the moment by defining the hard
state to have an orbit averaged ASM count rate of $<$45\,cps (see
Sect.~\ref{sec:bknpower}) and the soft state to have an ASM count rate
$\ge$80\,cps.  Using these definitions, for the total time interval
from 1996--2004, Cyg~X-1 spent 75\% of the time in the hard state, 4\%
in the soft state and 21\% in between. Before MJD\,51300, the source
was in the hard state for 90\% of the time, in the soft state 4\% of
the time, and in between for $\sim$6\% of the time. These numbers
change dramatically after MJD\,51300, where the time spent in the hard
state decreased to 66\%, while the soft and intermediate phases
account for 34\% of all ASM data points. While these numbers are not
too precise because of our arbitrary cuts between the different states
in terms of the ASM count rate, they clearly point towards a change in
source behavior in recent years, which might have been foreshadowed by
a change in the timing properties of the source around MJD~50920 (see
Fig.~3 of paper~\textsc{i}). Specifically, around this time the
short time scale rms variability (i.e., the rms associated with
individual pointed observations) significantly decreased, while the
long time scale variabilty (i.e., the prevalence of failed state
transitions) increased.

The physical reason for this change of behavior after $\sim$MJD 51000
is unclear. In its simplest interpretation, the soft state occurs at
higher $\dot{M}$, and thus one could speculate that the mass accretion
rate increased around that date.  As $\dot{M}$ is related to the mass
loss rate of HDE~226868, in principle it is possible to measure
$\dot{M}$ from its correlation with the equivalent width (EW) of the
H$\alpha$ line \citep{puls:96a}.  The simplest picture, that a higher
mass loss rate of HDE~226868 results in a higher $\dot{M}$ for the
black hole, however, is contradicted by the observations. Contrary to
expectations, the soft state H$\alpha$ EW is generally lower than in
the hard state
\citep{voloshina:97a,brocksopp:99b,gies:02a,tarasov:03a}.  For single
stars, such a behavior indicates a \emph{lower} mass loss rate
\citep[][and therein]{gies:02a}. A possible solution to this enigma is
that photoionization by the X-ray source changes the wind ionization
state \citep{hatchett:77a}. Observationally, this effect could be the
cause of the variation of the H$\alpha$ line profile with orbital
phase \citep{gies:02a,tarasov:03a}. To explain the relation between
the X-rays and the optical data, both teams of authors suggest that a
decrease in the mass loss rate of HDE~226868, $\dot{M}_\text{w}$,
changes the properties of the accretion flow onto the black hole. A
decreased wind density results in an increase of the size of the
Str\"omgren sphere around the X-ray source, resulting in a lower
radiative acceleration and consequential lower terminal wind velocity,
$v_\text{w}$. Since the mass accretion rate is
$\dot{M}\propto\dot{M}_\text{w}/v_\text{w}^4$
\citep{bondi:44a,davidson:73a}, an increase of $\dot{M}$ through the
accretion disk is triggered, leading to the increase in soft X-ray
emission.  The sparse optical coverage prohibits so far direct tests
whether this picture is true, although at least for one failed state
transition, \citet{tarasov:03a} find a decrease of the EW before the
flare, consistent with the above model.  If this general idea is
correct, it then follows that the recent increase in soft state
activity of Cyg~X-1 could be linked to episodes of lower mass loss
from HDE~226868 \citep{gies:02a}, e.g., linked to activity cycles of
the donor star.

\begin{figure*}
\includegraphics[width=12cm]{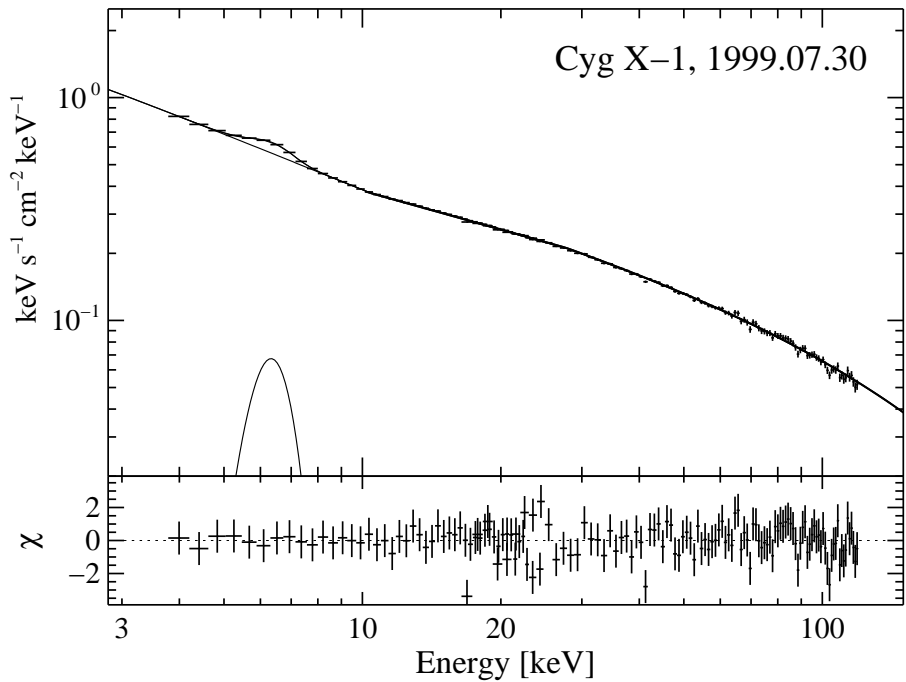}
\hfill\begin{minipage}[b]{5.5cm}
\caption{Unfolded spectrum and residuals for the Cyg X-1 observation
  of 1999 July 30 (P40090/14), using the broken power law model as the
  baseline continuum.  \label{fig:bknpower1999.07.30}}
\end{minipage}
\end{figure*}

\subsection{Choice of Spectral Models}
A major issue with describing Galactic black hole X-ray spectra is the
choice of the models to be employed; a multitude of more or less
physically motivated models are available. It is often found that
several of these models describe the X-ray data equally well in a
$\chi^2$ sense \citep{nowak:02a}. Unfortunately, however, it can be
difficult to directly compare spectral parameters obtained from these
physically motivated models, as they often make different implicit
assumptions in their physical setup \citep[][and references
therein]{nowak:02a,coppi:04a}. For example, Comptonization models
differ in their assumptions on the spectral shape of the seed photons
(e.g., Wien spectrum versus black body versus accretion disk spectrum)
and they also assume different geometries for the source of the seed
photons and of the Comptonizing medium \citep{nowak:02a,coppi:04a}.
These assumptions greatly complicate a direct comparison even of
``simple'' parameters such as the covering factor of the reflecting
medium.

Here, we use an approach already used by us earlier in a study of
GX~339$-$4 \citep{nowak:02a} to describe the spectral evolution of
Cyg~X-1 using several different spectral models. We start in
Sect.~\ref{sec:bknpower} with the most simple spectral model, a broken
power law with an exponential cutoff. This model gives a very good
empirical description of the hard state X-ray spectral shape. We then
use two different physically motivated models to find an
interpretation of the empirically derived spectral shape in terms of
Comptonization. In Sect.~\ref{sec:comptt} we use the widely available
Comptonization model of \citet[see also
\citealt{tit:95a,tit:95b}]{tit:94a}, \texttt{compTT}. We contrast the
\texttt{compTT} results in Sect.~\ref{sec:eqpair} with the more
elaborate Comptonization model \texttt{eqpair}
\citep{coppi:92a,coppi:99a}, which provides a good example for todays
class of self consistent Comptonization models that also include the
capability to generate the spectra resulting from non-thermal
Comptonization.

\subsection{Empirical models: Broken Power Law Fits}\label{sec:bknpower}
To describe the spectrum of Cyg X-1 on purely empirical grounds, as
originally motivated by our modeling of the broad band radio to X-ray
spectrum of GX~339$-$4 and Cyg~X-1, \citep{nowak:05a}, we model the
joint PCA/HEXTE data with an absorbed broken power law with break
energy $E_\text{break}$, soft photon index $\Gamma_1$, and hard photon
index $\Gamma_2$, that is exponentially cutoff above an energy
$E_\text{cut}$ with an $e$-folding energy of $E_\text{fold}$. In
addition to this continuum, all fits require an Fe K$\alpha$ line at
$\sim$6.4\,keV, which we model as a Gaussian with energy $E_\text{Fe
  K$\alpha$}$ and width $\sigma_\text{Fe K$\alpha$}$. Table~1,
available in electronic form only (see footnote~\ref{fn:aatab}),
contains all of these best fit parameters and their 90\% confidence
intervals, including $\Gamma_1$, $\Gamma_2$, $E_\text{fold}$,
$E_\text{break}$, and the Fe line parameters. The table also includes
photon and energy fluxes determined from the model for the 2--5\,keV,
5--10\,keV, 10--50\,keV, 50--100\,keV bands and the unabsorbed
2--100\,keV flux. Furthermore, the table includes the time of each
observation, its exposure time, and the PCA and HEXTE count rates.

Fig.~\ref{fig:bknpower1999.07.30} shows a typical example for a broken
power law fit, illustrating that for most observations this continuum
model gives a good description of the data, with
$\chi^2_\text{red}<2$.  A similar result for broken power law fits has
been found previously, e.g., by \citet{gierlinski:99a}.  In general,
we find that during the hard state the goodness of the broken power
law fits as measured in terms of their $\chi^2_\text{red}$ surpasses
that of the \texttt{compTT} and \texttt{eqpair} fits. During the soft
state, the broken power law fits still provide a very good description
of the hard spectrum above $\gtrsim$6\,keV, but fare less well below
$\sim$6\,keV where the accretion disk becomes important\footnote{ This
  failure of the simple broken power law is most evident for
  observations P50110/53 (2002.05.06:00), P60090/05 (2002.05.06:02),
  P60090/07 (2002.05.31:23) and P60090/15 (2002.09.20:16).  These data
  are characterized by $\Gamma_1\gtrsim 3$, i.e., they are amongst the
  softest in our sample and there is a clear thermal component
  present.  As our main interest in this section is on the behavior of
  the harder spectrum, we ignored the data below 5\,keV for these two
  observations.  Above that energy, the spectrum could again be well
  described by a broken power law.  Additionally, there are two
  peculiar observations, P50110/07 (2000.06.02:05) and P50110/10
  (2000.07.15:00). In these observations, $\Gamma_1\sim \Gamma_2$ (but
  not equal!) and when both power law indices and the Fe line
  parameters are left free, the uncertainties of both power law
  indices are very large, the break energy in these observations
  converges to $\sim$5.5\,keV and the energy of the Gaussian is found
  close to 5.5\,keV with a width in excess of 1\,keV, effectively
  smoothing out the effect of the break.  Rather than forcing the
  power law indices to some (arbitrary) value, we have decided to omit
  these two observations from the following analysis, although they
  are contained in Table~1.}.  For these data, \texttt{compTT} and
\texttt{eqpair} give a better description of the spectrum.

\begin{figure*}
\centering

\includegraphics[width=88mm]{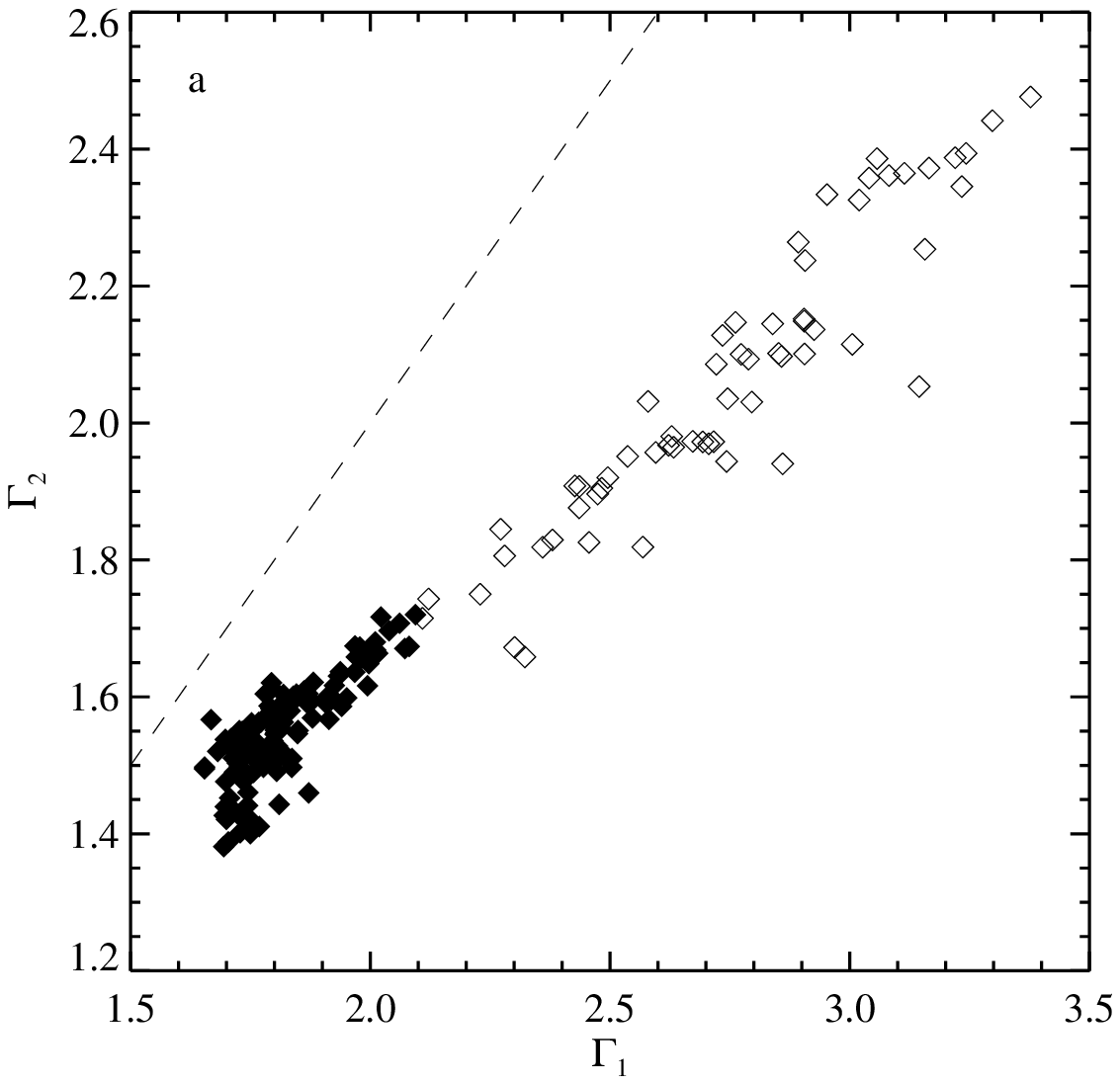}
\hfill
\includegraphics[width=88mm]{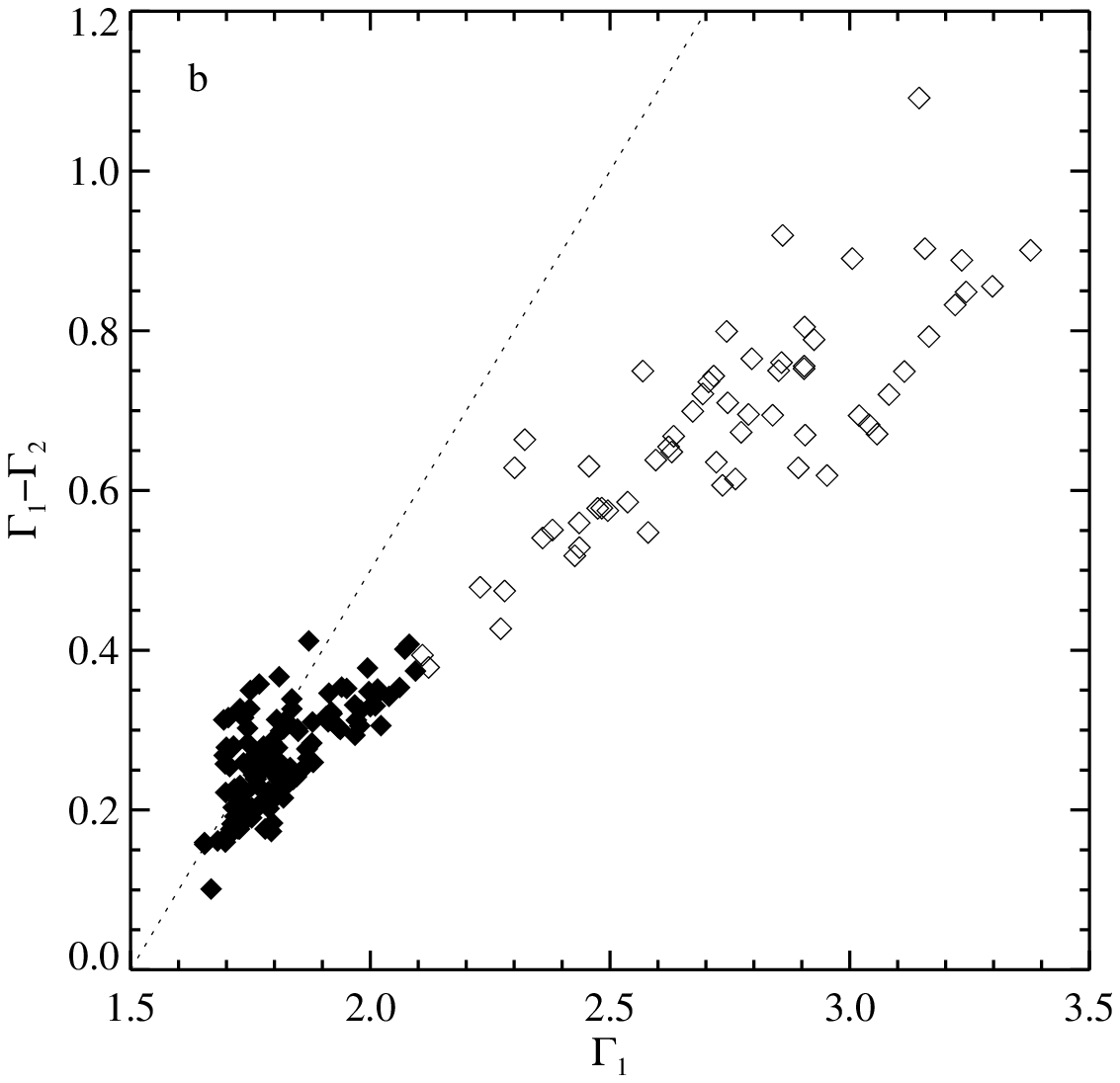}

\caption{\textbf{a)} Correlation between the soft power law index,
  $\Gamma_1$, and the hard power law index, $\Gamma_2$, for the broken
  power law fits. The dashed line corresponds to $\Gamma_1=\Gamma_2$.
  Filled diamonds designate hard state observations, defined by
  $\Gamma_1\le 2.1$, open diamonds are intermediate and soft state
  observations.  \textbf{b)} Correlation between soft energy power law
  index, $\Gamma_1$, and power law break, $\Gamma_1-\Gamma_2$, for the
  broken power law fits. Lines parallel to the dotted line corrspond
  to lines of constant $\Gamma_2$. For both correlations, Spearman's
  rank correlation coefficient $\rho=0.916$.}\label{fig:deltagamma}
\end{figure*}

\begin{figure}
\includegraphics[width=88mm]{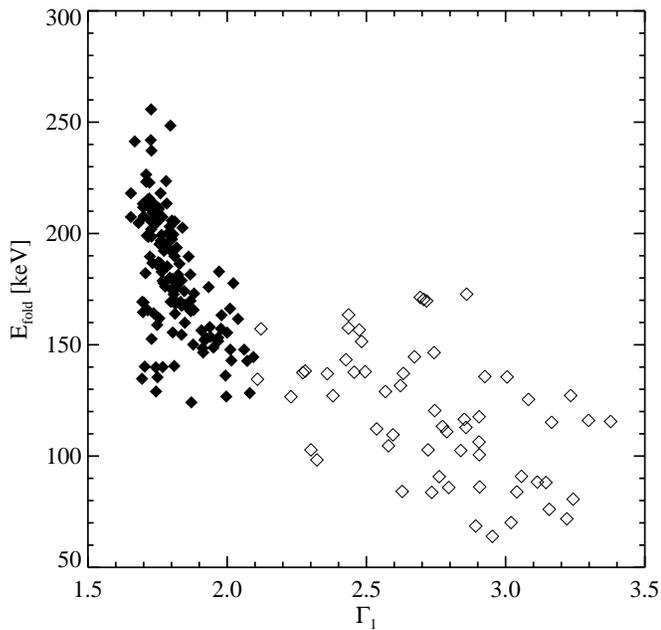}
\caption{Correlation between the photon index of the soft power
  law index, $\Gamma_1$, and the $e$-folding energy of the hard
  continuum, $E_\text{fold}$, for the broken power law
  fits. Symbols are the same as in
  Fig.~\ref{fig:deltagamma}}\label{fig:gamma1_efold} 
\end{figure}

As shown in Fig.~\ref{fig:deltagamma}a, for all observations of
Cyg~X-1 studied here, $\Gamma_2<\Gamma_1$, that is, the higher energy
power law is spectrally harder than the soft one. For the majority of
our fits, the break energy of the power laws is found between 15 and
40\,keV. There is a weak correlation between $E_\text{break}$ and
$E_\text{fold}$ in the sense that higher break energies also have
higher folding energies. A similar correlation exists between
$\Gamma_1$ and $E_\text{fold}$ (Fig.~\ref{fig:gamma1_efold}), where we
find that a softer continuum implies more curvature at high energies
than harder spectra. In Fig.~\ref{fig:deltagamma} and all following
figures showing spectral parameters, we show data points from the hard
state with the filled symbolds and data from observations from the
intermediate and soft state using open symbols.  To define the hard
state, we use $\Gamma_1<2.1$ \citep[a simplified version of the
classification of][]{remillard:05a}, corresponding to ASM count
rates below 45\,$\text{counts}\,\text{s}^{-1}$.

The most interesting property of the broken power law fits is the
hardening of the broken power law, which can be physically interpreted
as the hardening of the underlying continuum caused by Compton
reflection off cold or mildly ionized material.  As has been pointed
out by \citet{zdz:99a}, for Seyfert galaxies and galactic black hole
binaries, there is a correlation between the power law index and the
covering factor of the reflecting medium, $\Omega/2\pi$, in the sense
that softer power laws tend to show stronger reflection.  Our broken
power law fits show a similar correlation: There is a strong linear
relationship between $\Gamma_1$ and $\Gamma_2$
(Fig.~\ref{fig:deltagamma}a).  Correlating the soft power law index,
$\Gamma_1$ with the hardening of the broken power law at the break
energy, $\Delta\Gamma=\Gamma_1-\Gamma_2$, results in a similarly tight
correlation (Fig.~\ref{fig:deltagamma}b).  As discussed, e.g., by
\citet[and references therein]{zdz:99a}, for standard power law plus
reflection fits, $\Gamma$ and $\Omega/2\pi$ are not strictly
independent fit parameters: During $\chi^2$ minimization, softer power
law continua can be compensated to some extent by increasing
$\Omega/2\pi$. This behavior can lead to artificial correlations
between these two parameters. While the range of $\Omega$ and $\Gamma$
used to derive the observed $\Omega$--$\Gamma$ correlation was much
larger than the systematic uncertainty caused by the non-independent
fit parameters \citep{zdz:99a}, for sources where $\Omega$ and
$\Gamma$ do not vary as much, the artificial correlation is still of
some concern. In contrast, in the broken power-law fits $\Gamma_1$ and
$\Gamma_2$ are virtually independent and therefore artificial
correlations are not an issue for our analysis.

Lastly, we comment on the behavior of the Fe K$\alpha$ line in the
broken power law fits. In general, we find the line at 6.4\,keV, i.e.,
consistent with neutral iron, although spectra showing a larger
$\Delta\Gamma$ tend to show slightly higher line energies. In no case
does the line energy exceed 6.7\,keV. This trend could indicate a
slightly higher degree of disk ionization for the softer spectra
during these episodes.  The distribution of the line width,
$\sigma_\text{Fe K$\alpha$}$, from all observations can be well
described by a Gaussian distribution with a mean of 0.6\,keV,
consistent, e.g., with earlier \textsl{BeppoSAX} and \textsl{RXTE}
results \citep{frontera:01a,gilfanov:99b}. We performed Monte Carlo
simulations, confirming our experience that with its 18\% FWHM energy
resolution at 5.9\,keV the PCA can resolve lines down to $\sigma\sim
0.3$\,keV. The fits indicate, therefore, that the Fe K$\alpha$ line is
slightly broadened, as has been found also with higher resolution
instruments \citep[e.g.,][]{miller:02a}. Note, however, that the
$\sim$5\,keV calibration feature discussed in
Appendix~\ref{app:speccal} as well as the presence of a soft excess
can bias the line parameters towards a lower energy and larger
$\sigma$.

\subsection{Simple Comptonization Models: \texttt{compTT}}\label{sec:comptt}
We now turn to modeling the data with more physically motivated
spectral models. We first use the Comptonization model \texttt{compTT}
\citep{tit:94a,tit:95a,tit:95b}, as historically it has met with good
success in describing black hole spectra.  Model parameters discussed
here, therefore, can be used in comparisons with earlier results and
other sources \citep[see, e.g.,][]{pottschmidt:03a}. Furthermore, as
we shall see below, the behavior of \texttt{compTT} parameters is very
similar to that of the more sophisticated Comptonization models, while
being substantially faster to fit.

We took \texttt{compTT} as the baseline continuum, assuming a disk
geometry. The parameters of the continuum are the electron optical
depth, $\tau_\text{e}$, and the electron temperature $kT_\text{e}$. An
accretion disk continuum with a $r^{-3/4}$ temperature profile and
inner disk temperature $kT_\text{in}$ (XSPEC model \texttt{diskbb};
\citealt{mitsuda:84a,makishima:86a}) was added to model the soft
emission.  We set the temperature of the seed photons for
Comptonization equal to $kT_\text{in}$, indicating that the disk is
the source of the seed photons. Note that this approach is only
approximately justified, since the seed photon distribution assumed by
\texttt{compTT} is a Wien distribution and not a proper disk spectrum.
We take Compton reflection of the Comptonization continuum into
account using the Green's functions of \citet[XSPEC model
\texttt{reflect}]{magdziarz:95a}, assuming solar abundances, a neutral
reflector, an inclination of $40^\circ$, and taking into account
absorption in the interstellar medium and the stellar wind.  The
behavior of the latter is similar to that seen in the \texttt{eqpair}
fits, so we will defer a discussion of the behavior of the Hydrogen
equivalent column, $N_\text{H}$, to Sect.~\ref{sec:eqpair}. All best
fit parameters, including $kT_\text{in}$, $\tau$, $kT_\text{e}$,
$\Omega/2\pi$, the Fe line parameters, and the model normalizations,
as well as their 90\% confidence intervals are available in the
electronic Table~2 (see footnote~\ref{fn:aatab}). The table also
includes fluxes from the model in the same standard bands as Table~1,
as well as the total bolometric flux inferred from the model. Finally,
the table also includes the total bolometric accretion disk flux.

The Fe K$\alpha$ line was again modeled by a Gaussian, but with the Fe
K$\alpha$ line energy fixed at 6.4\,keV (fits with the line energy
free were virtually identical).  There is one clear difference between
the Fe K$\alpha$ parameters in the broken power law fits and the ones
found with \texttt{compTT}: Fe K$\alpha$ line widths are greater by a
factor of $\sim$1.55 than in the broken power law fits.  Inspection of
the \texttt{compTT} residuals shows the PCA to have a wavy structure
at low energies around the Fe line, indicating that our choice of soft
spectral components does not adequately describe the details of the
shape of the soft continuum.  In addition, we see the K$\alpha$ line
width increasing almost linearly with the accretion disk flux, until
it saturates at $F_\text{disk}\sim
10^{-8}\,\text{erg}\,\text{cm}^{-1}\,\text{s}^{-1}$ with a line width
of $\sim$1\,keV. Such a behavior is similar to that of a similarly
broad line feature (at $\sim$5.7\,keV) seen in \object{XTE~J1908+094},
in the early and late phases of the outburst of this soft X-ray
transient \citep{gogus:04a}.  The broader Fe K$\alpha$ line width can
thus be seen as an attempt of the $\chi^2$-minimization procedure to
smooth out the continuum to provide a more power-law like soft
continuum.  This conclusion is verified by the generally worse
$\chi^2$ of the \texttt{compTT} fits compared to the broken power law
ones, although in general the model is still providing a good
description of the data: For hard state and intermediate state
observations, $\chi^2_\text{red}\lesssim 1.3$, for soft state
observations, $\chi^2_\text{red}$ is clearly higher, although (except
for four cases) still better than 2.  Note that only in 12 out of the
202 observations is the $\chi^2_\text{red}$ of the \texttt{compTT}
fits better than that of the broken power law fits.

Replacing the disk component with a relativistic disk spectrum (XSPEC
model \texttt{diskpn}, \citealt{gierlinski:99a}) resulted in
significantly worse fits.  Adding a second soft component (modeled
either by a second disk spectrum or by optically thick Comptonization
\citealt{gierlinski:99a}) to our baseline \texttt{diskbb} model did
not result in improving the residuals significantly either. We therefore
 decided to stick with our baseline model, especially since the
Comptonization parameters are mainly driven by the harder spectrum.

\begin{figure}
\includegraphics[width=88mm]{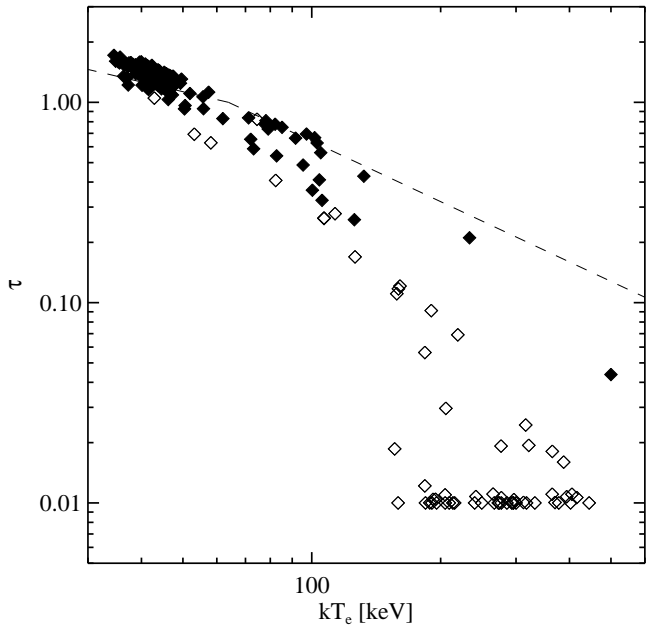}
\caption{Correlation between the electron temperature of
  the Comptonizing plasma, $kT_\text{e}$ and its optical depth,
  $\tau$. The dashed line shows the values for which the Compton $y$
  parameter $y=0.5$, the lower limit for $\tau$ set in the fits was
  $\tau=0.01$. Symbols used are identical to
  Fig.~\ref{fig:deltagamma}.}\label{fig:kt_taup}
\end{figure}

Fig.~\ref{fig:kt_taup} shows the correlation between the optical
depth, $\tau$, and the temperature $kT_\text{e}$ of the Comptonizing
plasma. Defining the Compton $y$ parameter in the usual fashion,
\begin{equation}
y=\frac{ 4 kT_\text{e}}{m_\text{e} c^2} \max\left(\tau,\tau^2\right)
\end{equation}
we find that for a large fraction of the data ($\sim$70\% of the
observations), $y\simeq 0.5$ (Fig.~\ref{fig:kt_taup}, dashed line).
That is, during these times the Compton-$y$ of the Comptonizing plasma
is nearly constant to within our error bars. This behavior is a well
known property of plasmas held in thermal equilibrium by balancing the
available heating by Compton cooling
\citep[e.g.,][]{dove:97a,dove:97b}. Note that we set the lower limit
of $\tau$ to 0.01 in our fits. For $\tau\lesssim 0.1$, if $y$ were to
remain constant at such low optical depths, the plasma would have
relativistic temperatures, violating the approximations used in the
\texttt{compTT} model \citep{tit:95b}.  Processes such as
photon-photon pair production would become important. For these
conditions, more realistic Comptonization models such as
\texttt{eqpair} have to be used (Sect.~\ref{sec:eqpair}). We still
find that $y$ is well determined for these observations, with
observations with $kT_\text{e}\gtrsim 100$\,keV leaving the track of
constant $y$ inferred above as $\tau$ pegs at its lower limit.
Virtually all of these observations were made during soft state
episodes and are therefore characterized by a strong soft excess and a
softer hard spectrum than the average hard state spectrum: The
majority of spectra with a hard power law index of $\Gamma_2>1.8$ in
the broken power law fits come from observations for which the
\texttt{compTT} fits yield $kT_\text{e}>100$\,keV. This result is
consistent with earlier studies of the 1996 soft state of Cygnus~X-1
\citep[e.g.,][]{cui:96a}.

\begin{figure}
\includegraphics[width=88mm]{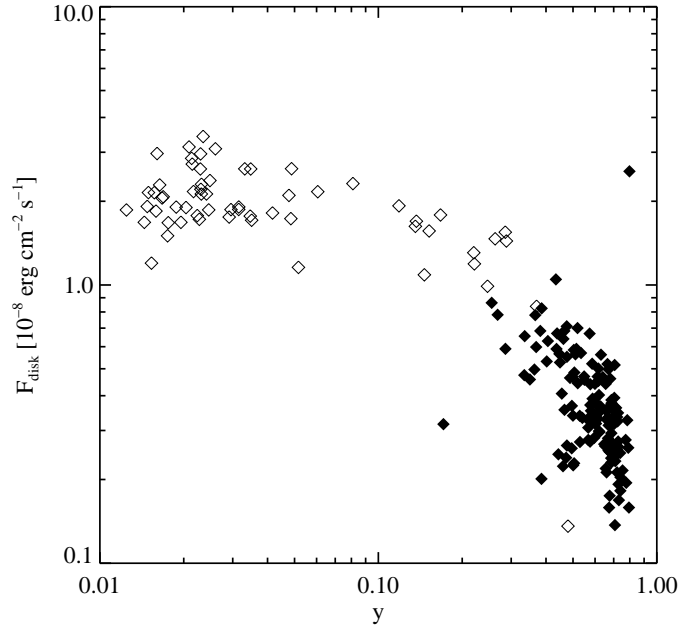}
\caption{Relationship between Compton-$y$  and the unabsorbed accretion
  disk flux for the \texttt{compTT} fits. Symbols used are the same as
  in Fig.~\ref{fig:deltagamma}.}\label{fig:y_diskflux}
\end{figure}

\begin{figure}
\includegraphics[width=88mm]{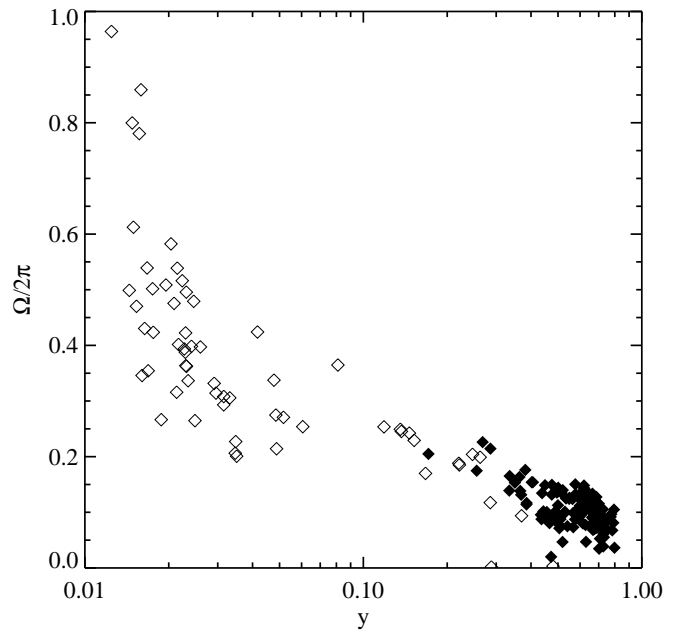}
\caption{Correlation between $y$
  and the covering factor for reflection, $\Omega/2\pi$ for the
  \texttt{compTT} fits, showing that $\Omega$ strongly increases for
  small $y$. Symbols used are the same as in
  Fig.~\ref{fig:deltagamma}.}\label{fig:y_omega}
\end{figure}

Since the parameter responsible for the shape of the Comptonization
spectrum is the Compton-$y$, we now study the relationship between the
accretion disk and the reflection parameters and $y$
(Figs.~\ref{fig:y_diskflux} and~\ref{fig:y_omega}). In general, the
correlations found in this way are much tighter than correlations with
respect to $kT_\text{e}$ and $\tau$, as these two parameters are
strongly coupled.

We compute the bolometric unabsorbed flux of the accretion disk,
$F_\text{disk}$, by integrating the \texttt{diskbb} component of the
best fit model from 1\,eV to 50\,keV. The second source of soft
photons, the Wien seed photon spectrum, is not taken into account.
Comparison with the soft fluxes derived with \texttt{eqpair}, where
the seed photons can be taken into account (see Sect.~\ref{sec:eqpair}
below), shows that not taking the Wien spectrum into account results
in the bolometric disk fluxes from \texttt{compTT} to be 73\% of the
\texttt{eqpair} fluxes. After correcting the \texttt{compTT} fluxes
for this offset, the standard deviation of the the \texttt{eqpair} and
\texttt{compTT} ratio is found to 20\%, which we take as the
systematic uncertainty of the individual fluxes caused by
extrapolating the best fit model outside of the energy range used for
fitting. This 20\% systematic uncertainty of individual flux points is
much smaller than the overall variation of the disk fluxes found in
the data. Note that in the following, when quoting $F_\text{disk}$
from \texttt{compTT} fits we do \emph{not} apply any correction
factor, such as might be caused by ignoring the Wien seed spectrum.

\begin{figure*}
\includegraphics[width=12cm]{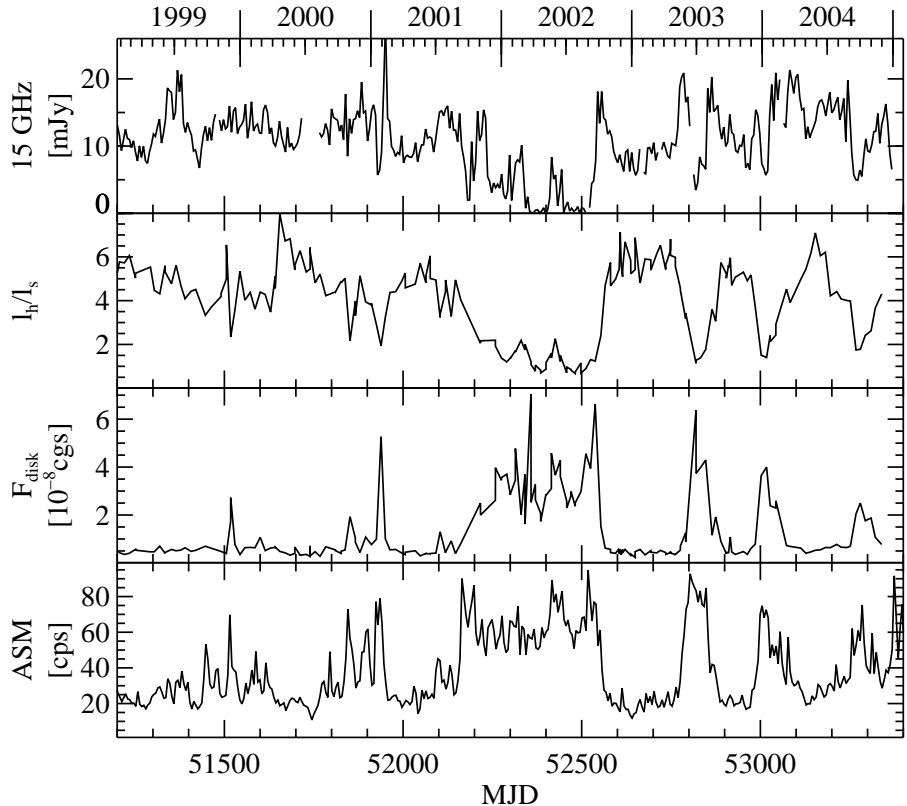}\
\hfill\begin{minipage}[b]{5.5cm}
\caption{Spectral evolution of Cyg X-1 from 1999 until the end of
  2004. Shown from top to bottom are the radio flux at 15\,GHz as
  measured with the Ryle telescope, the compactness ratio, $\lhls$ and
  the bolometric flux of the accretion disk as obtained from the
  \texttt{eqpair} models, and the 2--10\,keV \textsl{RXTE}-ASM light
  curve, rebinned to a resolution of 5.6\,d to smooth out the orbital
  variability. Error bars are not shown for
  clarity.}\label{fig:eqp_evol_radio}
\end{minipage}
\end{figure*}

Fig.~\ref{fig:y_diskflux} shows how $F_\text{disk}$ depends on $y$.
There are clearly two regimes of different spectral behavior
recognizable: During the hard state (filled symbols), $F_\text{disk}$
is seen to vary by almost an order of magnitude, while $y$ remains
constant. At the same time, $\Omega/2\pi$ varies only slightly
(Fig.~\ref{fig:y_omega}). This behavior of parameters holds for disk
fluxes below
$\sim$$10^{-8}\,\text{erg}\,\text{cm}^{-2}\,\text{s}^{-1}$,
corresponding to a threshold disk luminosity of $4.8\times
10^{36}\,\text{erg}\,\text{s}^{-2}$ \citep[assuming a distance of
2\,kpc, see][and references therein]{ziolkowski:05a}.  Above this
threshold the intermediate or the soft state is reached.  Here, $y$ is
seen to vary strongly, the disk luminosity is far less variable than
before, and $\Omega/2\pi$ varies strongly.

\subsection{Comptonization: eqpair}\label{sec:eqpair}
We turn to the most sophisticated of the spectral models discussed
here, the \texttt{eqpair} model \citep{coppi:92a,coppi:99a}.  It is a
hybrid thermal/nonthermal Comptonization code, and it includes
electron-positron pair production.  Given the overall success of
thermal Comptonization to describe the hard X-ray spectrum described
in Sect.~\ref{sec:comptt}, in the following we assume a fully thermal
electron distribution.  \texttt{Eqpair} computes the temperature of
the Comptonizing medium self-consistently, by balancing external
heating with Compton cooling of a defined seed-photon distribution,
which in our case is the \texttt{diskpn} model \citep{gierlinski:99a}.
The amount of heating of the Comptonizing medium is specified in
\texttt{eqpair} through the ratio of the compactnesses of the
Comptonizing medium and the seed photon distribution, $\lhls$, where
the dimensionless compactness parameter of a medium of characteristic
size $r$ is defined by
\begin{equation}
\ell=\frac{L \sigma_\text{T}}{r m_\text{e} c^3}
\end{equation}
where $L$ is the source luminosity, $\sigma_\text{T}$ the Thomson
cross section, $m_\text{e}$ the electron mass, and $c$ the speed of
light. To force the seed photon spectrum to be dominated by disk
radiation, we fix $\ell_\text{s}=10$.  A fraction of the hard X-rays
is scattered back onto the disk where it is reflected. The amount of
reflection is again quantified by the covering factor for reflection,
$\Omega/2\pi$, contrary to the \texttt{compTT} fits, relativistic
smearing is taken into account for the reflection component.

Similar to the \texttt{compTT} fits, we model the soft excess by
adding a \texttt{diskbb} to the continuum. In contrast, detectors with
a lower energy threshold than the PCA often have shown the soft X-ray
spectrum in \texttt{eqpair} modeling to be more complex than what we
use here.  The soft excess has been described by optically thick
Comptonization of 100--300\,eV disk photons \citep{balu:95a}, as
predicted for the inner regions of \citet{shakura:73a} disks
\citep{gierlinski:97a,frontera:01a,disalvo:01a}. Our spectra have an
$\sim$3\,keV lower threshold, thus we cannot describe in detail this
component.  Similar to the \texttt{compTT} fits, we find that adding a
\texttt{diskbb} spectrum to the \texttt{eqpair} continuum results in a
satisfactory description of the data.  We set the inner disk
temperature to the maximum temperature of the \texttt{eqpair} seed
photon distribution.  Modeling the soft excess with the
\texttt{diskbb} works better than with the \texttt{diskpn}, as the
shape of the \texttt{diskbb} model better approximates the shape of a
(saturated) Comptonization spectrum.

Absorption in the interstellar medium and in the stellar wind is taken
into account using the model of \citet{balu:92a}, letting the hydrogen
equivalent column vary freely. We find a dependence of $N_\text{H}$
with orbital phase \citep[using the ephemeris of][]{brocksopp:99a},
with $N_\text{H}$ being higher by a factor $\sim$2 during superior
conjunction, pointing towards significant absorption in the stellar
wind of HDE~226868 (see \citealt{wen:99a} for a similar conclusion
based on \textsl{RXTE}-ASM data). We also confirm earlier results by
\citet{balu:00a} of a secondary maximum of $N_\text{H}$ around orbital
phase 0.6, which earlier has been identified with absorption in the
accretion stream. No other fit parameters show a significant
dependence on orbital phase.

Using this approach, our fits gave $\chi_\text{red}^2$ values similar
to the models discussed in Sects.~\ref{sec:bknpower}
and~\ref{sec:comptt}, with only a very weak dependence of
$\chi^2_\text{red}$ on the source state (softer spectra tend to have
slightly worse $\chi^2_\text{red}$).  During the hard state,
$\chi^2_\text{red}$ for the \texttt{eqpair} fits is generally worse
than that of the broken power law fits, while it is better during the
soft state (we attribute the latter to us not including a thermal
component in the broken power law fits). All best-fit parameters,
including $\ell_\text{h}/\ell_\text{s}$, $\Omega/2\pi$, and the Fe
line parameters, and X-ray fluxes for the same bands as for the broken
powerlaw and \texttt{compTT} fits, as well as unabsorbed fluxes for
the accretion disk and for the broadband spectrum are contained in
Table~3, which is available in electronic form only (see
footnote~\ref{fn:aatab}).

Fig.~\ref{fig:eqp_evol_radio} shows the evolution of energetics of the
soft and the hard spectral components. Similar to the \texttt{compTT}
fits, we characterize the soft flux by the unabsorbed bolometric
accretion disk flux, $F_\text{disk}$.  This flux is computed as the
sum of the 1\,eV--50\,keV fluxes of the \texttt{diskpn}\footnote{The
  \texttt{eqpair} model has a toggle that allows one to calculate the
  flux of solely the seed photon component}. and \texttt{diskbb}
spectral components of the best fit models. As discussed in
Sect.~\ref{sec:comptt}, the systematic uncertainty of these fluxes is
estimated to be 20\%, much smaller than the variations observed.  Not
unexpectedly, the disk flux is correlated with the binary orbit
averaged \textsl{RXTE}-ASM count rate. A more careful analysis of the
lightcurve reveals, however, that there the ASM does not always fully
track $F_\text{disk}$. For example, $F_\text{disk}$ increases
significantly during the second and third of the three dominant soft
flares around MJD~51500 and remains weak during the first. We will
revisit these differences, which might be due to hysteresis effects
similar to BHC transient outbursts, below.  Overall, however, the
correlation between $F_\text{disk}$ and the ASM count rate is good,
with the Spearman rank correlation coefficient indicating a highly
significant correlation ($\rho=0.851$).  A linear regression gives
\begin{equation}\label{eq:fdisk_asm}
\frac{F_\text{disk}}{10^{-8}\,\text{erg}\,\text{cm}^{-2}\,\text{s}} =
0.0592 \frac{F_\text{ASM}}{\text{counts}\,\text{s}^{-1}} - 0.809
\end{equation}
where $F_\text{ASM}$ is the rebinned ASM count rate. Due to the
hysteresis effects mentioned above, we estimate the uncertainty of
$F_\text{disk}$ computed from Eq.~\ref{eq:fdisk_asm} to
$\sim$$5\,10^{-9}\,\text{erg}\,\text{cm}^{-1}\,\text{s}^{-1}$. We note
that evolution of the soft excess obtained from the \texttt{compTT}
fits is very similar to that from the \texttt{eqpair} fits, although
in general \texttt{compTT} gives a disk flux that is $\sim$70\% of
that found with \texttt{eqpair}.

\begin{figure}
\includegraphics[width=88mm]{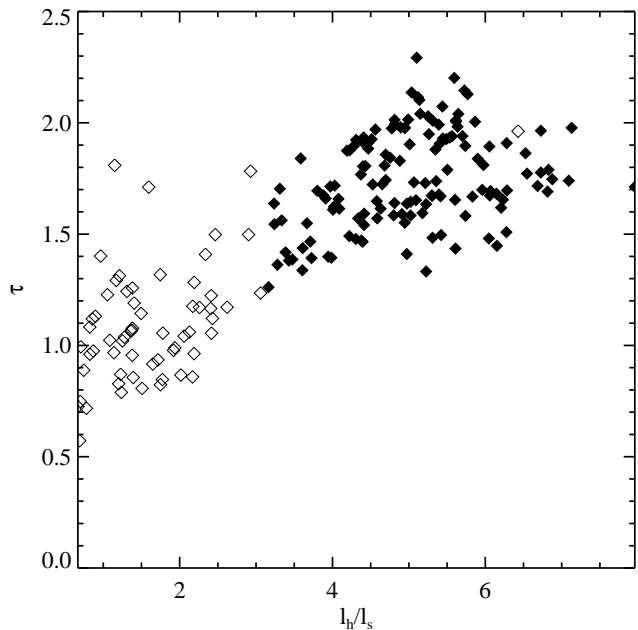}
\caption{Relationship between the compactness ratio of the
  Comptonizing medium, $\lhls$, and the optical depth $\tau$ of the
  Comptonizing medium for the \texttt{eqpair} models. Symbols are the
  same as in Fig.~\ref{fig:deltagamma}.}\label{fig:lhls_tau}
\end{figure}

\begin{figure}
\includegraphics[width=88mm]{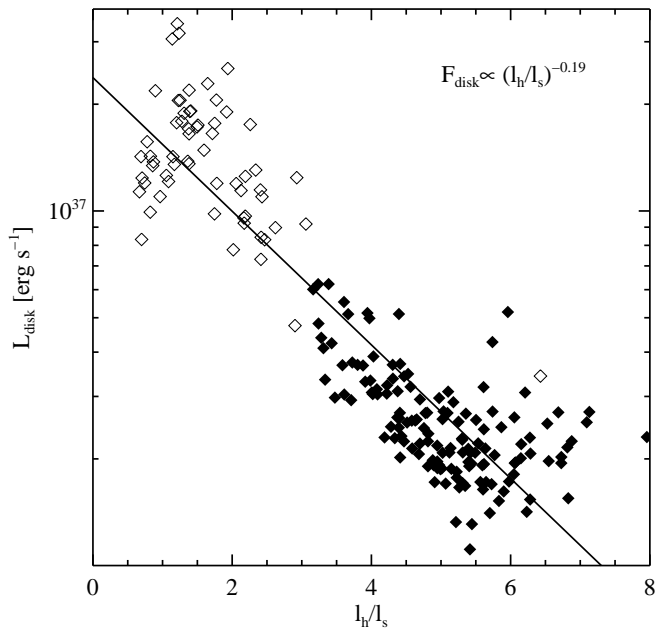}
\caption{The bolometric luminosity of the accretion disk,
  $L_\text{disk}$, computed assuming a distance of $d=2$\,kpc, shows a
  power law dependence on the compactness ratio of the Comptonizing
  medium, $\lhls$. Symbols used are identical to
  Fig.~\ref{fig:deltagamma}.}\label{fig:diskflux_lhls}
\end{figure}

\begin{figure*}
\includegraphics[width=88mm]{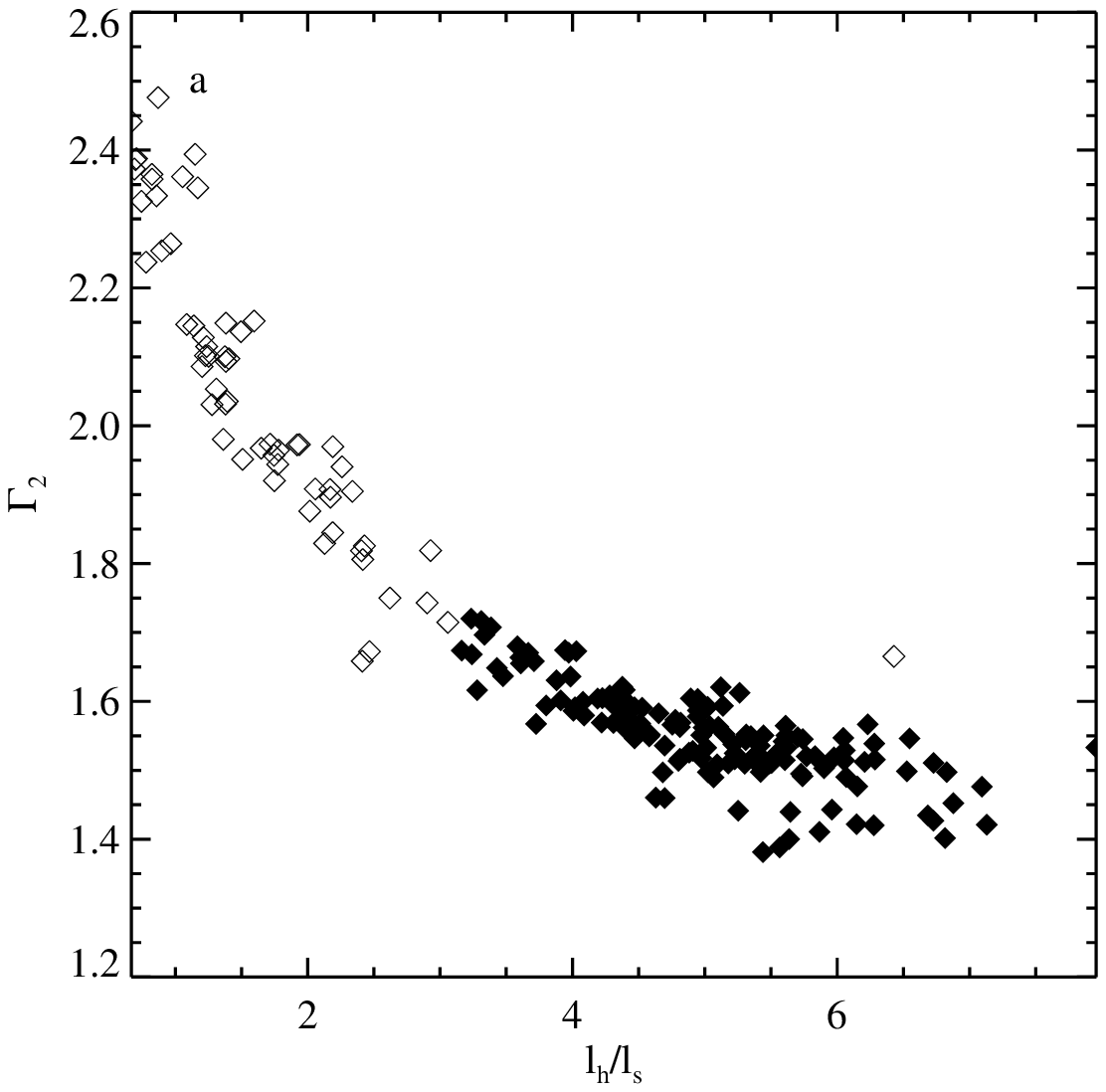}
\hfill
\includegraphics[width=88mm]{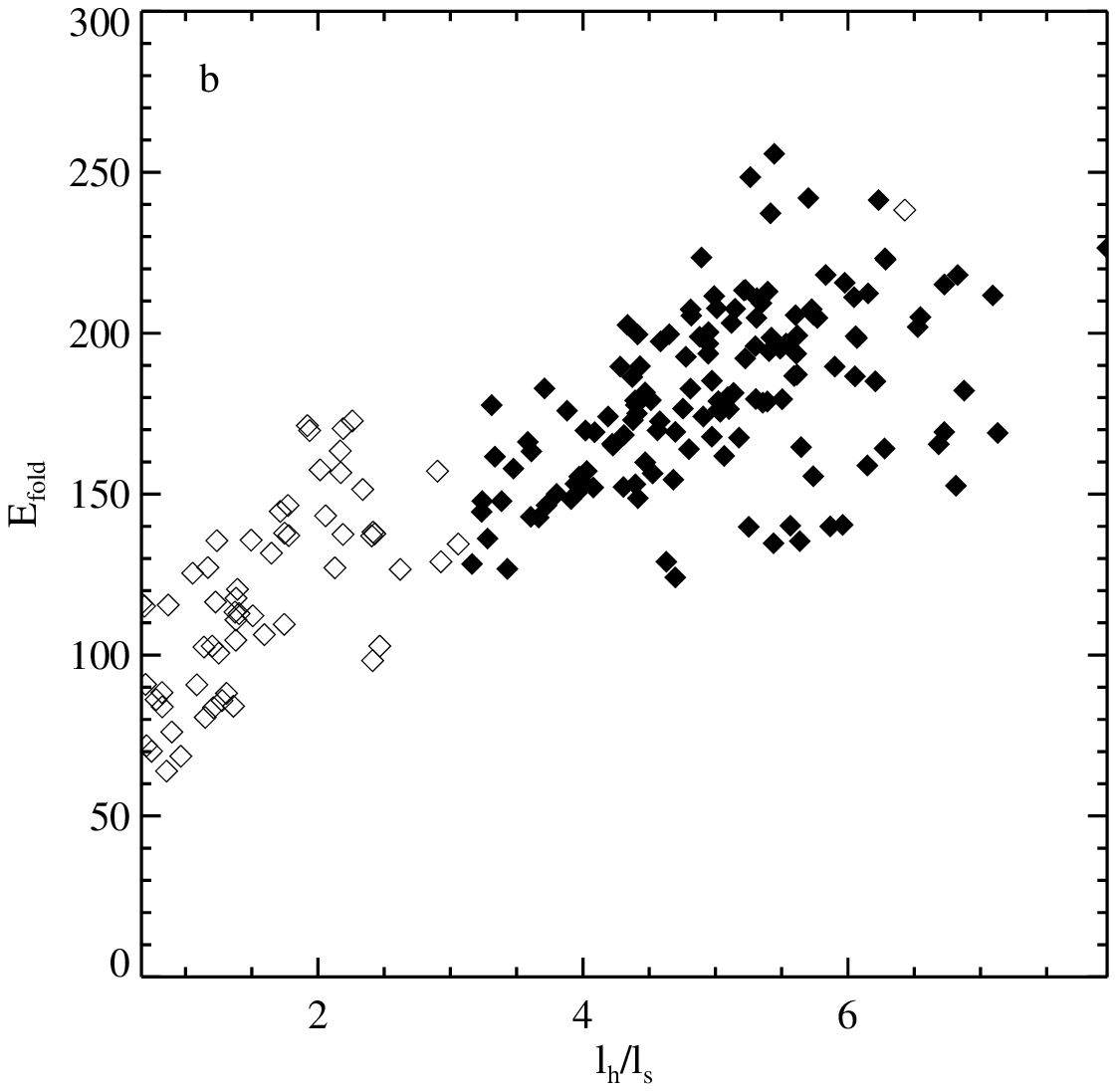}
\caption{Relationship between $\lhls$ and the shape of the X-ray
  spectrum above 10\,keV as found from the broken power law
  fits. \textbf{a)} Correlation between $\lhls$ and the photon index,
  $\Gamma_2$, \textbf{b)} Correlation between $\lhls$ and the folding
  energy, $E_\text{fold}$.  }\label{fig:eqpair_bkn} 
\end{figure*}

The hard spectral component is characterized by the compactness ratio,
$\lhls$ and the optical depth $\tau$ of the Comptonizing medium.
Large $\lhls$ correspond to the hard state. As shown in
Fig.~\ref{fig:lhls_tau}, $\tau$ increases with $\lhls$, indicating the
increased importance of the Comptonizing medium for higher $\lhls$.
Consequently, low $\lhls$ correspond to the soft state, such that
there is an anticorrelation between $\lhls$ and the accretion disk
luminosity (Fig.~\ref{fig:diskflux_lhls}).  This anticorrelation can
be expressed as a power law dependence of the bolometric accretion
disk luminosity, $L_\text{disk}$, and $\lhls$,
\begin{equation}
L_\text{disk} \propto (\ell_\text{h}/\ell_\text{s})^{-0.19}
\end{equation}
This relationship holds over the $\sim$1 order of magnitude in
variation of the disk luminosity.  Our fits thus indicate very smooth
relationships between the parameters of the Comptonizing medium and
the accretion disk, with no jumps being present that would indicate
that the source behavior suddenly changes when the source switches its
state. This behavior is also clearly seen in
Fig.~\ref{fig:eqpair_bkn}, where we show how the \texttt{eqpair}
parameters relate to the continuum shape as described by the broken
power law models. For large values of $\lhls$, the photon index
saturates at $\sim$1.6 and the folding energy approaches its canonical
value of $\gtrsim$150\,keV. Not unexpectedly, for lower $\lhls$ as the
soft state is approached, $\Gamma_2$ softens and $E_\text{fold}$
decreases, resulting in a more curved hard X-ray spectrum.

\begin{figure*}
\includegraphics[width=12cm]{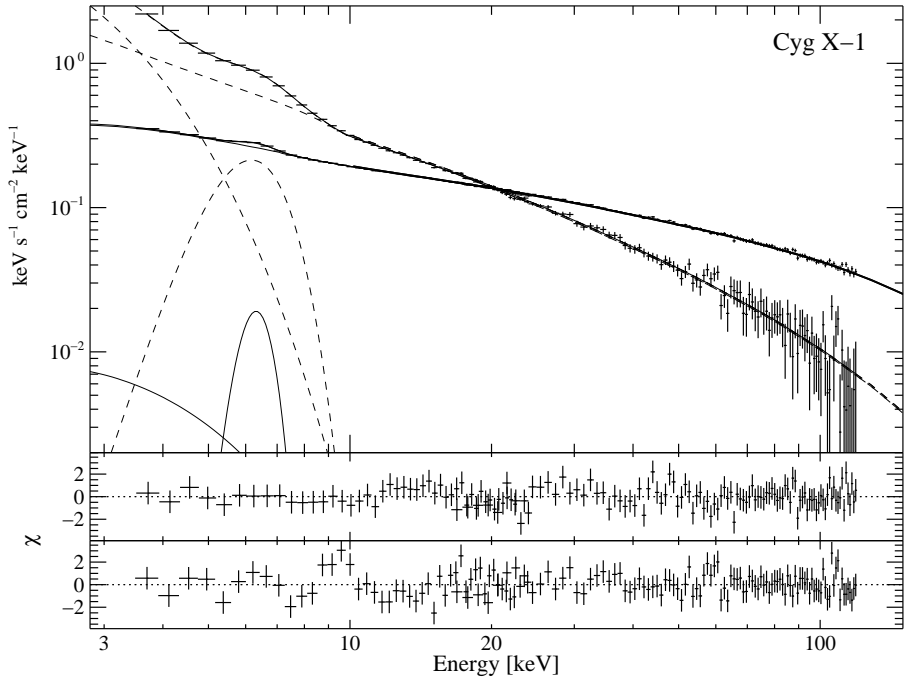}
\hfill\begin{minipage}[b]{5.5cm}
\caption{Unfolded spectra and residuals for the eqpair fits of the
  Cyg~X-1 observations of 2003 January 10 (P60090/23.14off; hard
  spectrum) and 2003 July 29 (P60090/35.14off; soft spectrum),
  illustrating the typical spectral variability of the source.
  \label{fig:eqpair_extremes}}
\end{minipage}
\end{figure*}

We now briefly turn to the evolution of the bolometric flux of
Cyg~X-1.  Based on \textsl{RXTE} and \textsl{CGRO} data,
\citet{zhang:97b} claimed that the difference in bolometric luminosity
of Cyg X-1 between the 1996 hard and soft states was $\lesssim
50\%$--70\%, a statement later revisited by \citet{zdz:02a}. Based on
a small number of observations, these authors found that the
bolometric flux of the source was higher by a factor of 3--4 in the
soft state. The availability of our broad band fits allows us to
confirm \citet{zdz:02a}'s earlier results: During the \textsl{RXTE}
campaign, the mean bolometric unabsorbed flux of Cyg~X-1 was
$4.3\,10^{-8}\,\text{erg}\,\text{cm}^{-2}\,\text{s}^{-1}$, with a
minimum flux of
$2.4\,10^{-8}\,\text{erg}\,\text{cm}^{-2}\,\text{s}^{-1}$ measured on
2003 January 10 (observation P60090/23.14off), and a maximum flux of
$9.1\,10^{-8}\,\text{erg}\,\text{cm}^{-2}\,\text{s}^{-1}$ measured on
2002 March 25 (observation P60090/02.14off). Assuming a distance of
2\,kpc, these values correspond to a mean source luminosity of
$2.1\,10^{37}\,\text{erg}\,\text{s}^{-1}$, varying between
$1.2\,10^{37}\,\text{erg}\,\text{s}^{-1}$ and
$4.4\,10^{37}\,\text{erg}\,\text{s}^{-1}$. These luminosity values are
extremes; in general we find the luminosity of Cyg~X-1 during the soft
states to be approximately twice that of the hard state phases.
Cyg~X-1 thus has a long-term average luminosity of
$(0.01$--$0.02)\,L_\text{Edd}$, assuming a black hole mass between 10
and 15 solar masses \citep{ziolkowski:05a,herrero:95a}.
Fig.~\ref{fig:eqpair_extremes} shows unfolded spectra for two of the
more extreme observations of Cyg X-1, illustrating the strong spectral
variability of the source.

\begin{figure}
\includegraphics[width=88mm]{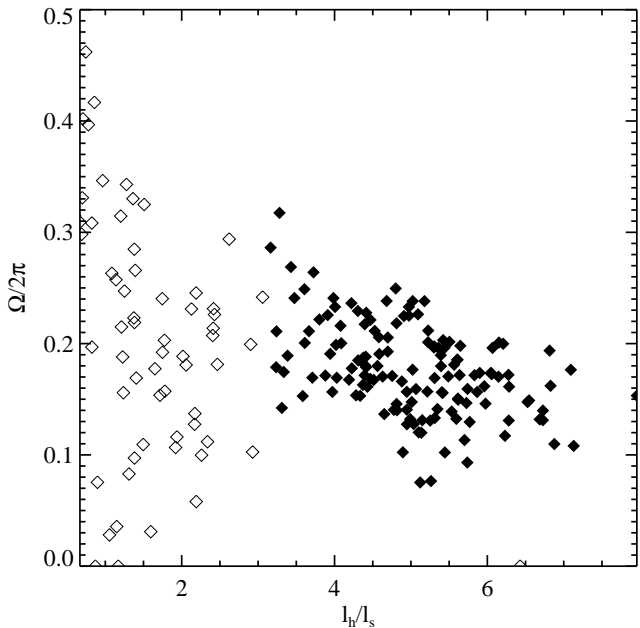}
\caption{Correlation between the compactness ratio of the Comptonizing
  medium, $\ell_\text{h}/\ell_\text{s}$, and the covering factor of
  the reflector, $\Omega/2\pi$. Symbols are the same as in
  Fig.~\ref{fig:deltagamma}. }\label{fig:lhls_omega}
\end{figure}

Finally, we discuss the variability of the reprocessing features,
i.e., the covering factor of the reflecting medium, $\Omega/2\pi$, and
the Fe K$\alpha$ line. Fig.~\ref{fig:lhls_omega} shows that
$\Omega/2\pi\sim 0.15$ for $\lhls\gtrsim 4$, indicating moderate
amounts of reflection during the hard state.  There is a moderate
trend for $\Omega/2\pi$ to decrease with increasing $\lhls$ (and
increasing $\tau$), although this trend is somewhat weaker and less
distinct than previously claimed anti-correlations between reflection
fraction and spectral hardness \citep[][and references
therein]{zdz:04a,zdz:99a}.  For our observations, the \texttt{eqpair}
and the \texttt{compTT} fits show a similar relative behavior;
however, $\Omega/2\pi$ derived from \texttt{eqpair} is larger by
$\sim$0.1.  This slightly larger fitted covering factor is likely due
to the usage of a relativistic smeared reflection continuum in the
\texttt{eqpair} fits; reflection in the \texttt{compTT} model is
artificially reduced to minimize the sharp Fe edge in the reflection
model used with those latter fits.  There may also be influences from
the different shapes of the exponential cutoff in these two models.
In the \texttt{eqpair} models, as the source approaches the soft state
and $\lhls$ decreases, there is no clear correlation with
$\Omega/2\pi$.  The reflection fraction varies in a seemingly
independent manner, although in general our highest fitted reflection
fractions come from the soft state.  Again this is reminiscent of the
behavior already seen in the \texttt{compTT} fits
(Fig.~\ref{fig:y_omega}).

The second indicator for reprocessing in cold material, the Fe
K$\alpha$ line, does not show a correlated behavior with any of the
spectral parameters considered here. This lack of correlation is
puzzling. Since the Fe K$\alpha$ line is thought to be the result of
fluorescence in the accretion disk, we would expect the Fe line to
track $\Omega/2\pi$, i.e., the variation of the Compton reflection
component. Similar to our fits with the broken power law or the
\texttt{compTT} model, there is a larger fraction of fits in which the
Fe K$\alpha$ line has an energy that is significantly below 6.4\,keV.
The residuals for these observations often show the Xe L edge
calibration feature, indicating that calibration effects might lead at
least in part of the observations to spurious Fe K$\alpha$ line
parameters, although we cannot exclude other effects, such as a more
complex shape of the soft excess. As already indicated above, however,
neither the addition of a Compton component \citep{gierlinski:99a} nor
the substitution of the \texttt{diskbb} soft component by a
\texttt{diskpn} component result in a better description of the soft
excess and the Fe K$\alpha$ line.

\subsection{Disk--Jet Interaction}\label{sec:diskjet}
As we have seen in Fig.~\ref{fig:eqp_evol_radio}, in agreement with
studies of other sources, we find evidence for correlated radio and
X-ray variability.  Due to the comparably low luminosity variation of
a factor of $\sim$4, Cyg~X-1 is not an ideal source to study disk-jet
correlations across widely separated states, as occur in X-ray
transients that show factors of 100 to 1000 luminosity changes
\citep{fender:04a,fender:04b}. The proximity of Cyg~X-1 to the
threshold between the hard and soft states, however, allows us to
study this crucial transition interval in much greater detail than for
transients.

\begin{figure}
\includegraphics[width=88mm]{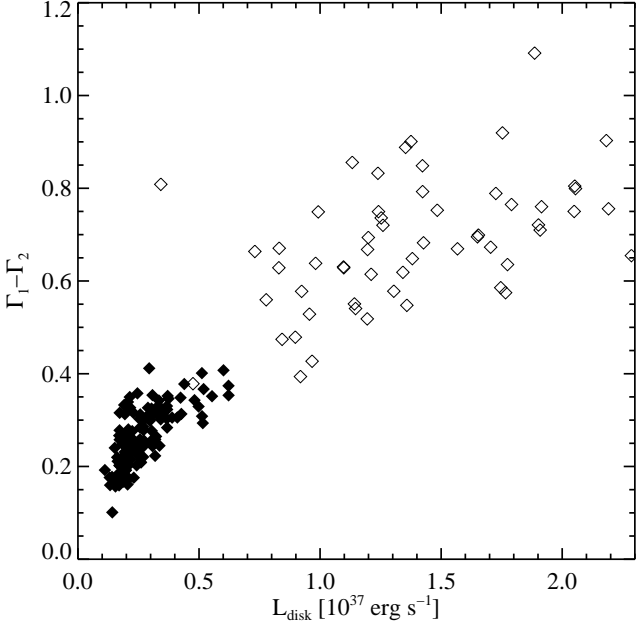}
\caption{The power law break, $\Delta\Gamma$, of the broken power law
  fits can be interpreted as a measure of the bolometric accretion
  disk luminosity.}\label{fig:diskflux_deltagamma}
\end{figure}

To parameterize our data with the least possible theoretical bias, we
can use the results from the broken power law fits. As shown in
Sect.~\ref{sec:bknpower}, these fits indicate the presence of two
spectral components that vary in a correlated manner as described by
the $\Gamma_1$--$\Gamma_2$ correlation. We show in
Fig.~\ref{fig:diskflux_deltagamma} that there is an almost linear
relation between the soft excess luminosity found from Comptonization
fits and $\Delta\Gamma=\Gamma_1-\Gamma_2$. In Comptonization models,
therefore, one can interpret the power law break in part as a measure
of the bolometric accretion disk luminosity, although part of the
break can also be caused by the $\Omega$-$\Gamma$-relationship.

\begin{figure}
\includegraphics[width=88mm]{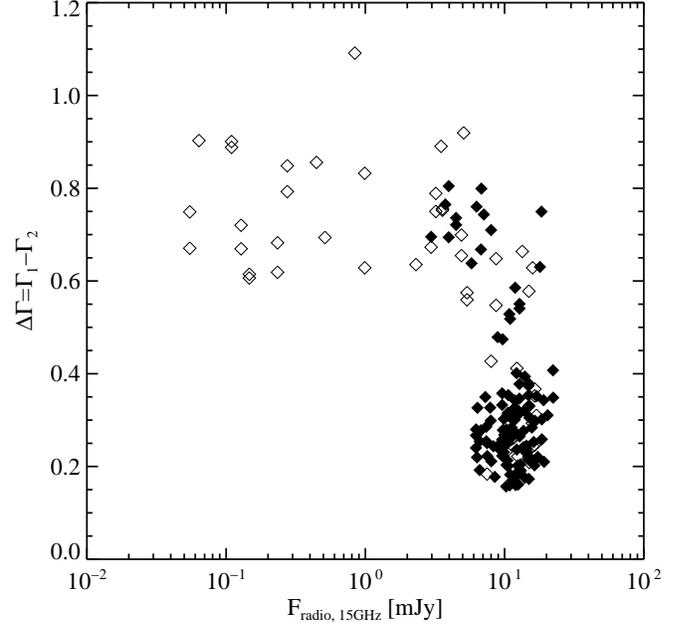}
\caption{For power law breaks $\lesssim$0.5 the radio flux varies
  seemingly independent of the X-ray color, expressed here as the
  ratio between the 2--10\,keV and the 10--50\,keV flux. Symbols used
  are identical to Fig.~\ref{fig:deltagamma}. See text for further
  explanation.}\label{fig:deltagamma_radio}
\end{figure}

Alternatively, however, in models explaining the X-ray spectra of BHC
in terms of emission from along a jet, the power law break describes
the transition from synchrotron radiation (as well as accretion disk
emission) to a synchrotron self-Compton spectrum that is modified by
Compton reflection
\citep{markoff:02a,markoff:05a,markoff:04a,nowak:05a}.
Figure~\ref{fig:deltagamma_radio} provides an important benchmark for
such models. For $\Delta\Gamma\lesssim 0.5$, i.e., most hard state
observations, the radio flux is seemingly independent of
$\Delta\Gamma$, while above that threshold strong variations of the
radio flux are observed.

\begin{figure}
\includegraphics[width=88mm]{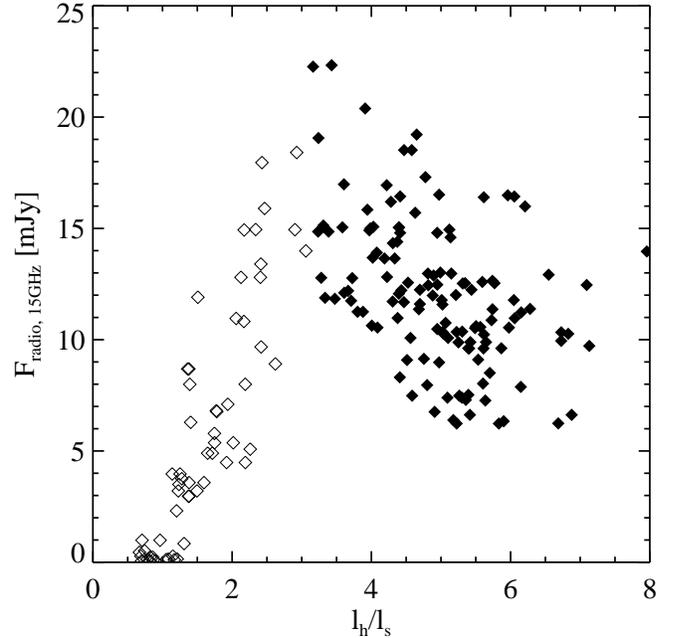}
\caption{Correlation between the 15\,GHz radio flux and
  the compactness ratio of the Comptonizing medium, $\lhls$. The
  increase in radio flux for intermediate values of $\lhls$ is due to
  the radio flares usually associated with ``failed state
  transitions'' and the intermediate state. The radio flux shown is
  the mean radio flux averaged over the 5.6\,d orbital timescale
  surrounding each \textsl{RXTE} measurement. Symbols used are the
  same as in Fig.~\ref{fig:deltagamma}.}\label{fig:lhls_radio}
\end{figure}

Using the Comptonization interpretation of the X-ray spectrum, we can
study how the radio emission and the Comptonizing medium interact. For
such a study the \texttt{eqpair} models are especially interesting,
since $\lhls$ is a measure of the energetics of the Comptonizing
medium.  We find that the radio flux is highest at intermediate
$\lhls$ ($\lhls\sim3$; Fig.~\ref{fig:lhls_radio}).  This behavior is a
consequence of the ``failed state transitions'', where Cyg~X-1 shows
characteristic flaring behavior in the radio, similar to what has also
been seen in GX~339$-$4 and other jet sources \citep[][see also
Figs.~\ref{fig:evolution}
and~\ref{fig:eqp_evol_radio}]{corbel:00a,corbel:03a,hannikainen:98a,fender:04b}.
Note that $\lhls=3$ corresponds to a soft photon index of
$\Gamma_1\sim 2.1$, consistent with the upper limit for the hard state
$\Gamma_1$ as defined by \citet{remillard:05a}. Note also that there
seems to be a much tighter relationship between the radio flux and
$\lhls$ for the soft state compared to the scatter seen for the hard
state.

\begin{figure}
\includegraphics[width=88mm]{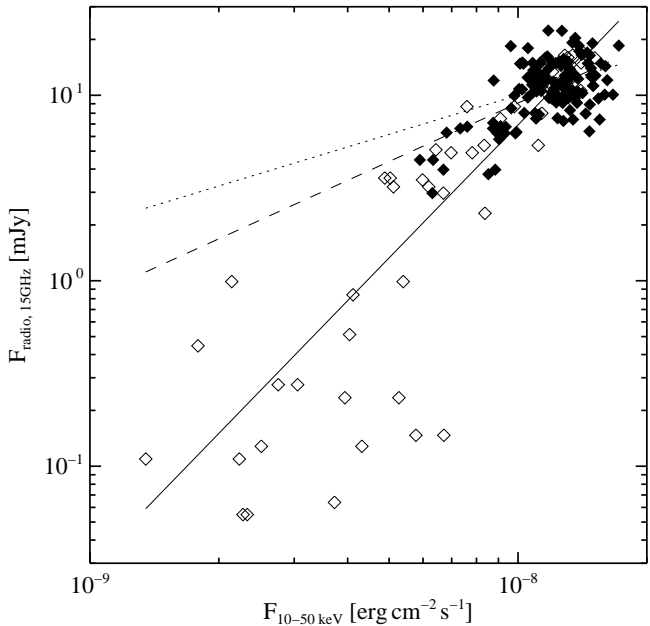}
\caption{Correlation between the 10--50\,keV hard X-ray flux and the
  15\,GHz radio flux. The dotted line corresponds to the canonical
  $F_\text{R}\propto F_\text{X}^{0.7}$ relation
  \citep{corbel:03a,gallo:03a}, the dashed line is $F_\text{R}\propto
  F_\text{X}^{1.05}$ determined from the hard state data, the solid
  line is $F_\text{R}\propto F_\text{X}^{2.38}$ found from all Cyg X-1
  measurements.}\label{fig:hardflux_radio}
\end{figure}

In general, BHC for which radio fluxes have been measured exhibit a
strong correlation between the radio and the X-ray flux
\citep{gallo:03a,merloni:03b}. For Cyg~X-1, this correlation becomes
less well pronounced as the soft X-ray flux increases
\citep[paper~\textsc{iii};][]{gallo:03a}.  We likewise do not find a
significant relationship between the radio flux and either the
bolometric flux or the disk emission. As already described in
paper~\textsc{iii} and by \citet{nowak:05a}, however, the radio flux
correlates well with the 10--50\,keV hard X-ray flux ($\rho=0.625$).
For 15\,GHz fluxes above 0.1\,mJy, the lowest flux at which the source
was detected during the campaign, we find that the 10--50\,keV X-ray
flux, $F_{10-50}$, can be found from
\begin{equation}\label{eq:radioflux}
\log_{10} \frac{F_{10-50}}{\text{erg}\,\text{cm}^2\,\text{s}^{-1}}
=
-8.27+0.320 \log_{10} \frac{F_{\text{15\,GHz}}}{\text{mJy}}
\end{equation}
with an uncertainty of less than 15\%. Here, the radio flux is again
the mean flux determined over one orbital period of the system to
average out the orbital variability \citep{brocksopp:99b}. 

This result is in strong support of the idea that the hard X-ray flux
is tightly coupled to the radio, suggesting a possible physical
connection between the regions in which this radiation is produced.
Note that for many X-ray transients the radio--X-ray relationship is
found from the 2--10\,keV \textsl{RXTE}-ASM data, which is possible
only because of the less complex soft excess in these Roche lobe
accreting systems.  Overall, however, the radio behavior of Cyg~X-1 as
expressed by Eq.~\ref{eq:radioflux} seems to be slightly different
from that of X-ray transients, where \citet{gallo:03a} and
\citet{corbel:03a} find that the radio flux, $F_\text{R}$, and X-ray
flux, $F_\text{X}$ are related by $F_\text{R}\propto
F_\text{X}^{0.7}$. Such a scaling is expected in various models in
which the X-ray and radio producing regions are connected, either by
the X-rays originating in a jet
\citep{falcke:95a,corbel:03a,markoff:02a,markoff:05a} or by the X-rays
originating in an advection dominated type of accretion flow that is
coupled to the radio outflow \citep{heinz:03a}.
Fig.~\ref{fig:hardflux_radio} shows that for Cyg~X-1 there are
indications that the radio--X-ray relationship is steeper than this
canonical value.  Taking all observations with available radio fluxes
into account, we find $F_\text{R}\propto F_\text{X}^{2.38}$
(Fig.~\ref{fig:hardflux_radio}, solid line).  Limiting ourselves to
the hard state observations only, where the radio fluxes are best
determined, we find $F_\text{R}\propto F_\text{X}^{1.05}$
(Fig.~\ref{fig:hardflux_radio}, dashed line), which, given the
uncertainty of the flux determination, can be considered as consistent
with the $F_\text{R}\propto F_\text{X}^{0.7}$ relationship. Similarly
steep radio--X-ray relationships have been seen in the black hole
candidate XTE J1908+094 \citep{jonker:04a} and in the neutron star
system \object{4U 1728$-$34} \citep{migliari:03a}.

\section{Summary}\label{sec:summary}
In this paper we have analyzed $\sim$200 observations of Cygnus~X-1 in
all of its states. The spectral parameters are in qualitative
agreement with those found for Comptonization models of earlier
observations
\citep[e.g.,][]{pottschmidt:03a,disalvo:01a,frontera:01a,gilfanov:99b,gierlinski:99a,dove:97c}.
The availability of the large number of observations allows us to
study the variation of the spectral parameters in this black hole
candidate in unprecedented detail. This has revealed a large number of
interesting correlations that must be explained by models for the
emission process and emission geometry of Cyg~X-1. In summary, the
main results of the spectral analysis are as follows:
\begin{enumerate}
\item Cyg~X-1 changed its behavior in early 2000. While it was
  predominantly a hard state source before that, since 2000 the source
  spent $\sim$34\% of the time in the intermediate and the soft state
  (see also paper~\textsc{i}). The physical reason for this change in
  behavior could be related to changes in the mass loss rate from
  HDE~226868.
\item The spectral continuum of Cyg~X-1 is very simple and can be well
  described at the level of the \textsl{RXTE} resolution by a broken
  power law with an exponential cutoff.
\item More physical models such as the two Comptonization models
  considered here try to reproduce this simplicity and, for the hard
  state, result in $\chi^2$ values comparable (but slightly worse)
  than the broken power law fits.
\item At the $\chi^2$ level, simple and advanced Comptonization models
  cannot be distinguished and the qualitative variation of their
  parameters is similar. From our modeling, a purely thermal electron
  distribution seems to be sufficient to describe the spectra in the
  hard and soft state observations, at least in the $\sim$3--120\,keV
  energy range considered here.
\item There is a continuum of spectra between the hard state and the
  soft state, with no strong discontinuities found in the behavior of
  the spectral parameters.  This continuous variation is in strong
  contrast with the changes in timing behavior, where the states are
  clearly distinguishable. For example, the intermediate state and
  failed state transitions have increased X-ray time lags
  \citep{pottschmidt:00b} and changed power spectra
  (paper~\textsc{i}).
\item This continuous distribution of parameters from the soft to the
  hard state leads to robust correlations between spectral parameters,
  reflecting trends in the continuum of spectral shapes of Cyg~X-1.
  For the broken power law fits, the most important correlation is
  that between the soft and the hard power law index, for the
  Comptonization fits there is a power law relationship between the
  soft excess luminosity and the compactness ratio, $L_\text{disk}
  \propto (\ell_\text{h}/\ell_\text{s})^{-0.19}$.
\item For the \texttt{compTT} models, the Comptonizing plasma shows a
  steady behavior during the hard state and strongly varies during the
  soft state (with a similar behavior in the reflection covering
  factor, $\Omega/2\pi$).  During the hard state, the soft excess
  properties vary seemingly independently from the Comptonizing
  plasma.
\item Although both Comptonization models show a slight
  anti-correlation between spectral hardening and reflection fraction,
  $\Omega/2\pi$, it is not as pronounced as has been previously
  claimed \citep[][and references therein]{zdz:04a}.  It is likely
  that part of this correlation is in fact related to correlations
  between broad-band soft and hard continuum components.
\item The Fe K$\alpha$ line is found to be only moderately broad in
  the broken power law fits, consistent with earlier results
  \citep{frontera:01a,gilfanov:99b}.  An increased broadening seen in
  the Comptonization models is likely not a physical effect but rather
  is caused by an attempt of the fits to improve the modeling of the
  soft excess component.
\item For the hard state, Cyg~X-1 follows the general radio--X-ray
  relation, $F_\text{R}\propto F_\text{X}^{0.7}$, although there are
  indications that the relationship is slightly steeper when also
  taking the soft state data into account.
\end{enumerate}
A further interpretation of these results in terms of physical models
requires the consideration of the simultaneous variation of the timing
parameters, which we will present in a further paper in this series.

\begin{acknowledgements}
  We thank Thomas Glei\ss{}ner, Sara Benlloch, and William ``Biff''
  Heindl for many discussions and help with the data screening and
  extraction during the years it took to assemble the data from the
  Cyg~X-1 campaign and the anonymous referee for his/her comments.
  This work has been partly funded by NASA contract NAS5-30720, NASA
  grant SV3-73016, by Deutsches Zentrum f\"ur Luft- und Raumfahrt
  grant 50\,OR\,302 and by travel funds from the Deutscher
  Aka\-demischer Austauschdienst and the National Science Foundation
  (NSF contract INT-0233441). The Ryle telescope is supported by
  PPARC. This work has made use of data obtained from the High Energy
  Astrophysics Science Archive Research Center (HEASARC), provided by
  NASA's Goddard Space Flight Center.  The Green Bank Interferometer
  was a facility of the National Science Foundation operated by the
  National Radio Astronomy Observatory in support of NASA High Energy
  Astrophysics programs.  The University of Warwick Centre for
  Scientific Computing provided significant computing resources
  through its Cluster of Workstations (COW). We thank the Aspen Center
  for Physics for its hospitality during the final stages of the
  preparation of this paper.
\end{acknowledgements}

\appendix

\section{Revisiting the \textsl{RXTE} spectral calibration}\label{app:speccal}

In this Appendix we summarize our results for the spectral calibration
of the RXTE based on HEASOFT 5.3.1.  As shown, e.g., by
\citet{wilms:98c}, earlier HEASOFT versions exhibited a significant
difference in spectral slope of $\sim$0.05 between the PCA and HEXTE,
with PCA spectra being consistently softer than spectra derived from
HEXTE and other missions. For observations of black holes, such a
difference is especially worrisome since the apparent hardening
implied by the older calibration can mimic a Compton reflection
component. In HEASOFT 5.3.1, this discrepancy is now much smaller and
fits to \textsl{RXTE} observations of the Crab nebula and pulsar give
the canonical spectral shape \citep[see][for an extensive discussion
of the PCA calibration]{jahoda:05a}. For example, modeling the joint
PCA and HEXTE data from \textsl{RXTE}'s Crab observation
40805-01-05-01 in 1999 we find that the joint spectrum can be
described as the sum of two power laws with photon indices
$\Gamma_1=1.45$ and $\Gamma_2=2.12$. The nebula flux at 1\,keV is
found to be
10.3\,$\text{photons}\,\text{cm}^{-2}\,\text{s}^{-1}\,\text{keV}^{-1}$,
or about 8\% higher than the canonical value of \citet{toor:74a}, the
hard pulsar continuum contributes
0.05\,$\text{photons}\,\text{cm}^{-2}\,\text{s}^{-1}\,\text{keV}^{-1}$
at this energy (all values given are unabsorbed fluxes, absorption in
the interstellar medium was taken into account by using the
cross-sections and abundances of \citealt{wilms:00c} and by assuming a
H-equivalent column of $N_\text{H}=4.24\times 10^{21}\,\text{cm}^{-2}$
as found with \textsl{Chandra} by \citealt{weisskopf:04a}).

\begin{figure}
\centering
\includegraphics[width=88mm]{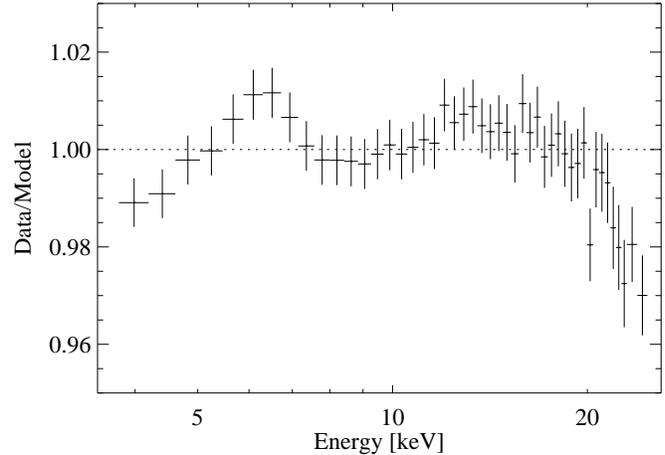}
\caption{Ratio between data and the two power-law spectral model for
 the joint PCA-HEXTE fit of the \textsl{RXTE} observation of the Crab nebula
 and pulsar 40805-01-05-01 described in the text. For clarity, only
 the PCA residuals are shown. The error bars reflect the 0.5\%
 systematics applied in our analysis.}\label{fig:crab}
\end{figure}
The ratio between the best fit Crab model (folded through the response
matrix) and the data reveals the presence of energy dependent
systematic uncertainties (Fig.~\ref{fig:crab}). These are strongest in
the 4--6\,keV band, around the Xe L-edge, where the deviation can
amount to up to 1\%, and above $\sim$25\,keV. We chose to take these
deviations into account by adding an energy independent uncertainty of
0.5\% to the data. Applying this uncertainty to the Crab data gives a
reduced chi-squared of $\chi^2_\text{red}=1.38$. While a larger
systematic uncertainty in the 4--6\,keV band and above $\sim$20\,keV
would be justifiable in light of this result, we have decided not to
include energy dependent systematics since experimenting showed that
such systematics would affect the Fe K$\alpha$ line region even more
strongly than our current strategy (where the Fe K$\alpha$ line energy
is already biased towards energies below 6.4\,keV). In addition, we
only use PCA data from channels 6--52 of the standard2f data mode
(approximately 3\,keV--25\,keV). For the HEXTE we use channels 17--117
(corresponding to $\sim$18\,keV to $\sim$120\,keV). The upper energy
bound of 120\,keV is set by the shorter observations in the sample and
by the desire to have an uniform energy range for all observations.
Note that HEXTE residuals in channels 17--20 typically show a linear
increasing trend, which might be due to calibration effects. We chose
to include these channels anyway to increase the overlap between the
PCA and HEXTE.  When ignoring these HEXTE channels, the $\chi^2$ of
our fits typically decreases without significant changes of the best
fit parameters.

In addition to the difference in spectral slope, earlier HEASOFT
versions also showed a very significant overestimate of the PCA flux
with respect to the HEXTE, with the PCA fluxes being consistently
higher by $\sim$30\%. This overestimate can be taken into account in
$\chi^2$-fitting by multiplying the spectral model instrument
dependent constant. Traditionally, this normalization constant was set
to 1 for the PCA and let vary for the HEXTE. This traditional approach
is unfortunate, since it implies that the flux normalization of the
HEXTE is uncertain, while in reality it was the PCA that showed clear
deviations in flux with respect to other instruments. For consistency
with other studies, however, here we continue using the traditional
approach. For all different spectral models used in this paper, we
find that under HEASOFT 5.3.1 both instruments now agree remarkably
well, with the distribution of the constant values being consistent
with a normal distribution of mean 1.01 and standard deviation 0.03.
We still chose to include the constant in our final fits, as in
individual observations a mismatch between the PCA and the HEXTE is
possible, for example for those observations where the PCA deadtime
becomes important\footnote{PCA spectra were not deadtime corrected as
  the spectral shape is not influenced by the deadtime correction and
  as good indicators for the PCA deadtime are not available for the
  time since the failure of the Xe layer of PCU0, which also changed
  the deadtime behavior of that PCU.}. There are only 7 observations
with constants more than 2$\sigma$ away from the mean. Except for the
significantly deviating normalization constant, these observations are
not remarkable in any other way, and are therefore included in our
further analysis, although the cause for the deviation is currently
unknown. We conclude, therefore, that for the overwhelming majority of
observations there is now agreement in the PCA and HEXTE derived
fluxes.

\section{Confidence Intervals and \textsl{RXTE} Systematic
  Errors}\label{app:systematics}

As shown in the previous appendix, the high signal to noise
observations from \textsl{RXTE} require us to take the systematic
error of the instrument into account in order to obtain reasonable
spectral parameters. In general, this is done by adding the systematic
error in quadrature to the Poisson errors estimated from the
observation. For instruments such as the PCA it can happen that the
systematic error starts dominating the overall error. As a
consequence, the standard
``$\chi^2_\text{min}+\Delta\chi^2$''-procedure for obtaining
confidence intervals, as outlined, e.g., by \citet{lampton:76a} or
\citet{bevington}, does not apply since it is based on the assumption
that the variance of the data points is due to a Gaussian
distribution. Since the addition of systematic errors decreases the
$\chi^2$ sum, confidence intervals estimated using the
\citet{lampton:76a} prescription in the presence of systematic errors
are significantly larger than what is justified from statistical
grounds.

\begin{table}
\caption{Error bars for a representative observation of Cyg X-1,
  computed using the prescription of \citet{lampton:76a}. See text for
  further explanation.}\label{tab:mcfake}
\begin{tabular}{llll}
\hline
\hline
Parameter & value & \multicolumn{2}{c}{90\% error bar} \\
          &       & \multicolumn{1}{c}{with} & \multicolumn{1}{c}{without} \\
          &       & \multicolumn{2}{c}{systematics}\\
\hline
$kT_\text{in}$ [keV]                & 0.855  & $(-0.02,+0.03)$  & $(-0.01,+0.06)$   \\
$A_\text{diskbb}$                   & 1276   & $(-175,+203)$    & $(-24,+40)$       \\
$kT_\text{e}$ [keV]                 & 108    & $(-31,+45)$      & $(-20,+27)$       \\
$\tau$                              & 0.26   & $(-0.14,+0.13)$  & $(-0.12,+0.15)$   \\
$A_\text{comptt}$  [$10^{-2}$]      & 1.86   & $(-1.0,+1.1)$    & $(-0.5,+0.7)$     \\
$\Omega/2\pi$  [\%]                 & 17.9   & $(-1.8,+1.3)$    & $(-1.1,+1.1)$     \\
$\sigma_\text{Fe K$\alpha$}$ [keV]  & 1.01   & $(-0.05,+0.04)$  & $(-0.006,+0.012)$ \\
$A_\text{Fe K$\alpha$}$ [$10^{-2}$] &  5.34  & $(-0.4,+0.4)$    & $(-0.07,+0.08)$   \\
$c_\text{HEXTE}$                    & 0.999  & $(-0.004,+0.005)$& $(-0.003,+0.003)$ \\
\hline
\end{tabular}
\end{table}

To study the influence of the systematic error on the confidence
intervals in our observations, we have used a Monte Carlo simulation:
We first determined the best fit parameters for the \texttt{compTT}
model of observation P60090/16 (2002.10.06:15) using the
\citet{lampton:76a} approach, including systematic errors. We then
used this best fit model to simulate how this observation would look
like if the PCA and HEXTE calibration were perfectly known. These
simulated spectra were then refit without applying a systematic error
and new confidence intervals were determined. The resulting 90\%
confidence intervals for both fits are shown in Table~\ref{tab:mcfake}
(Error bars at the 68\% level can be obtained from the 90\% confidence
intervals to a sufficiently high precision by assuming that the
uncertainty can be approximated by a normal distribution). Depending
on the parameter, the error bars are smaller by a factor of 2 to 5,
showing that the confidence interval is completely dominated by the
systematic error.

Since the systematic error will affect all observations of the Cyg~X-1
\textsl{RXTE} campaign in a similar way and since the major interest
is in determining \emph{trends} in the evolution of spectral
parameters, using the overestimation of the range of the confidence
intervals  determined from
individual observations would lead to misleading results. It is
better, therefore, to talk about an ``ensemble average'' in
determining the errors and to take the scatter found in individual
relationships between the fit parameters of Cyg~X-1 in a similar state
as an estimate their uncertainty. For this reason, we have decided to
not show the misleading individual error bars in the figures in the
main part of this paper, although they are available in the online
data accompanying this paper (Tables~1--3).
\end{document}